\RequirePackage{ifpdf}
\documentclass[nohyper,12pt,letterpaper]{JHEP3}
\usepackage{epsfig}
\usepackage[latin1]{inputenc}
\usepackage{bbm,amsfonts}
\usepackage{graphicx}
\usepackage{amssymb,amsmath}
\usepackage{fancybox,framed}
\usepackage{dsfont}
\usepackage{mathtools}

\author{Marco S. Bianchi$^{\bf a}$,
  Luca Griguolo$^{\bf b}$,
  Matias Leoni$^{\bf c}$,  
  Silvia Penati$^{\bf d}$
  and Domenico Seminara$^{\bf e}$\\
  $^{\bf a}$Institut f\"ur Physik,
Humboldt-Universit\"at zu Berlin,
Newtonstra{\ss}e 15, 12489 Berlin, Germany \\
  $^{\bf b}$ Dipartimento di Fisica e Scienze della Terra, Universit\`a di Parma and INFN Gruppo Collegato di Parma, Viale G.P. Usberti 7/A, 43100 Parma, Italy\\
  $^{\bf c}$  Physics Department, FCEyN-UBA \& IFIBA-CONICET\
Ciudad Universitaria, Pabell\'on I, 1428, Buenos Aires, Argentina \\
  $^{\bf d}$Dipartimento di Fisica, Universit\`a di Milano--Bicocca and
  INFN, Sezione di Milano--Bicocca, Piazza della Scienza 3, I-20126 Milano, Italy \\
  $^{\bf e}$ Dipartimento di Fisica, Universit\`a di Firenze and INFN Sezione di Firenze, via G. Sansone 1, 50019 Sesto Fiorentino, Italy   \\
  \qquad\\
  E-mail: \email{ marco.bianchi@physik.hu-berlin.de, luca.griguolo@pr.infn.it, 
  leoni@df.uba.ar, silvia.penati@mib.infn.it, seminara@fi.infn.it}
}

\abstract{We study a family of circular BPS Wilson loops in ${\cal N}=6$ super Chern--Simons--matter theories, generalizing the usual 1/2--BPS circle. The scalar and fermionic couplings depend on two deformation parameters and these operators can be considered as the ABJ(M) counterpart of the DGRT latitudes defined in ${\cal N}=4$ SYM. We perform a complete two--loop analysis of their vacuum expectation value, discuss the framing dependence and propose a general relation with cohomologically equivalent bosonic operators. We make an all--loop proposal for computing the Bremsstrahlung function associated to the 1/2--BPS cusp in terms of these generalized Wilson loops. When applied to our two--loop result it reproduces the known expression.  Finally, we comment on the generalization of this proposal to the bosonic 1/6--BPS case.}

\preprint{February 2014\\HU-EP-14/04}

\title{BPS Wilson loops and Bremsstrahlung function in ABJ(M): a two loop analysis}


\csname @addtoreset\endcsname{equation}{section}

\def\bseq{\begin{subequation}}  
\def\eseq{\end{subequation}}
\def\bsea{\begin{subeqnarray}}  
\def\esea{\end{subeqnarray}}

\newcommand{\beq}{\begin{equation}}
\newcommand{\bea}{\begin{eqnarray}}
\newcommand{\eea}{\end{eqnarray}}
\newcommand{\eeq}{\end{equation}}

\newcommand {\non}{\nonumber}

\renewcommand{\a}{\alpha}
\renewcommand{\b}{\beta}

\renewcommand{\d}{\delta}
\newcommand{\pa}{\partial}
\newcommand{\g}{\gamma}
\newcommand{\G}{\Gamma}

\newcommand{\e}{\epsilon}

\renewcommand{\l}{\lambda}
\renewcommand{\L}{\Lambda}

\newcommand{\p}{\pi}

\newcommand{\s}{\sigma}

\renewcommand{\t}{\tau}

\def\Mb{\kern 2pt\mathchoice
        {
         \vbox{\hrule width10pt height 0.4pt depth 0pt
         \kern 1.2pt\hbox{\kern -2pt$\displaystyle M$}}}
        {
         \vbox{\hrule width10pt height 0.4pt depth 0pt
         \kern 1.2pt\hbox{\kern -2pt$\textstyle M$}}}
        {
\vbox{\hrule width6pt height 0.4pt depth 0pt
         \kern 1.0pt\hbox{\kern -2pt$\scriptstyle M$}}}
        {
         \vbox{\hrule width5pt height 0.4pt depth 0pt
         \kern 0.8pt\hbox{\kern -2pt$\scriptscriptstyle M$}}}}

\def\Sb{\kern 2pt\mathchoice
        {
         \vbox{\hrule width6pt height 0.4pt depth 0pt
         \kern 1.2pt\hbox{\kern -2pt$\displaystyle S$}}}
        {
         \vbox{\hrule width6pt height 0.4pt depth 0pt
         \kern 1.2pt\hbox{\kern -2pt$\textstyle S$}}}
        {
         \vbox{\hrule width3.5pt height 0.4pt depth 0pt
         \kern 1.0pt\hbox{\kern -2pt$\scriptstyle S$}}}
        {
         \vbox{\hrule width3pt height 0.4pt depth 0pt
         \kern 0.8pt\hbox{\kern -2pt$\scriptscriptstyle S$}}}}

\def\Rb{\kern 2pt\mathchoice
        {
         \vbox{\hrule width5.5pt height 0.4pt depth 0pt
         \kern 1.2pt\hbox{\kern -2.5pt$\displaystyle R$}}}
        {
         \vbox{\hrule width5.5pt height 0.4pt depth 0pt
         \kern 1.2pt\hbox{\kern -2.5pt$\textstyle R$}}}
        {
         \vbox{\hrule width3.5pt height 0.4pt depth 0pt
         \kern 1.0pt\hbox{\kern -2.2pt$\scriptstyle R$}}}
        {
         \vbox{\hrule width3pt height 0.4pt depth 0pt
         \kern 0.8pt\hbox{\kern -2.2pt$\scriptscriptstyle R$}}}}

  \def\pp{{\mathchoice
      {
          \kern 1pt%
          \raise 1pt
          \vbox{\hrule width5pt height0.4pt depth0pt
            \kern -2pt
            \hbox{\kern 2.3pt
              \vrule width0.4pt height6pt depth0pt
              }
            \kern -2pt
            \hrule width5pt height0.4pt depth0pt}%
            \kern 1pt
       }
        {
          \kern 1pt%
          \raise 1pt
          \vbox{\hrule width4.3pt height0.4pt depth0pt
            \kern -1.8pt
            \hbox{\kern 1.95pt
              \vrule width0.4pt height5.4pt depth0pt
              }
            \kern -1.8pt
            \hrule width4.3pt height0.4pt depth0pt}%
            \kern 1pt
        }
        {
          \kern 0.5pt%
          \raise 1pt
          \vbox{\hrule width4.0pt height0.3pt depth0pt
            \kern -1.9pt  
            \hbox{\kern 1.85pt
              \vrule width0.3pt height5.7pt depth0pt
              }
            \kern -1.9pt
            \hrule width4.0pt height0.3pt depth0pt}%
            \kern 0.5pt
        }
        {
          \kern 0.5pt%
          \raise 1pt
          \vbox{\hrule width3.6pt height0.3pt depth0pt
            \kern -1.5pt
            \hbox{\kern 1.65pt
              \vrule width0.3pt height4.5pt depth0pt
              }
            \kern -1.5pt
            \hrule width3.6pt height0.3pt depth0pt}%
            \kern 0.5pt
        }
    }}

  \def\mm{{\mathchoice
               {
                 \kern 1pt
           \raise 1pt    \vbox{\hrule width5pt height0.4pt depth0pt
                  \kern 2pt
                  \hrule width5pt height0.4pt depth0pt}
                 \kern 1pt}
               {
                \kern 1pt
           \raise 1pt \vbox{\hrule width4.3pt height0.4pt depth0pt
                  \kern 1.8pt
                  \hrule width4.3pt height0.4pt depth0pt}
                 \kern 1pt}
               {
                \kern 0.5pt
           \raise 1pt
                \vbox{\hrule width4.0pt height0.3pt depth0pt
                  \kern 1.9pt
                  \hrule width4.0pt height0.3pt depth0pt}
                \kern 1pt}
               {
               \kern 0.5pt
         \raise 1pt  \vbox{\hrule width3.6pt height0.3pt depth0pt
                  \kern 1.5pt
                  \hrule width3.6pt height0.3pt depth0pt}
               \kern 0.5pt}
               }}

\def\pd{{\kern0.5pt
           + \kern-5.05pt \raise5.8pt\hbox{$\textstyle.$}\kern
0.5pt}}

\def\pmd{{\kern0.5pt
          \pm \kern-5.05pt
\raise6.3pt\hbox{$\textstyle.$}\kern1.5pt}}

\def\md{{\mathchoice
   {
      {{\kern 1pt - \kern-6.2pt \raise5pt\hbox{$\textstyle.$}\kern
1pt}}}
    {
      {{\kern 1pt - \kern-6.2pt \raise5pt\hbox{$\textstyle.$}\kern
1pt}}}
    {
      {\kern0.5pt - \kern-5.05pt
\raise3.4pt\hbox{$\textstyle.$}\kern0.5pt}}
    {
      {\kern0.5pt - \kern-5.05pt
\raise3.4pt\hbox{$\textstyle.$}\kern0.5pt}}}}

\def\beq{\begin{equation}}
\def\eeq{\end{equation}}
\def\bea{\begin{eqnarray}}
\def\eea{\end{eqnarray}}

\def\a{\alpha}
\def\b{\beta}
\def\g{\gamma}

\def\d{\delta}
\def\e{\epsilon}

\def\th{\theta}
\def\l{\lambda}
\def\G{\Gamma}

\def\L{\Lambda}

\newcommand{\Tr}{{\rm Tr}}
\newcommand{\Str}{{\rm Str}}

\newcommand{\facc}{\frac{\Gamma^2( \tfrac12-\epsilon)}{32 \pi^{3-2\epsilon}}(1-2\epsilon)\,}

\begin{document}

\section{Introduction and summary of the results}

In gauge theories Wilson loops are among the most important physical observables to be studied. In fact, since they are non--local operators, they encode information about the strong coupling regime of these theories. For instance, infinite Wilson lines provide the interaction potential between two heavy charged particles and allow for a consistent description of confinement in QCD. They also play a fundamental role at perturbative level and are at the very root of the lattice formulation. 

Remarkably, after the advent of the AdS/CFT correspondence, a new interest in Wilson loops for supersymmetric gauge theories has been triggered by their pivotal role in testing the correspondence itself. In fact, BPS Wilson loops are in general non--protected quantities and their vacuum expectation values undergo non--trivial flow between weak and strong coupling regimes. Therefore, whenever their vev is exactly computable, for instance summing the perturbative series or using localization techniques, they provide exact functions which interpolate from weak to strong coupling. This allows for non--trivial tests of the AdS/CFT predictions \cite{Maldacena}-\cite{Drukker:2000rr}. 

More recently, for ${\cal N}=4$ SYM, null--polygonal Wilson loops in twistor space have been proved to determine the exact expression for all--loop scattering amplitudes in the planar limit \cite{oai:arXiv.org:1009.2225}. At the same time, important duality relations between Wilson loops and scattering amplitudes have been found both at weak and strong coupling, which have been crucial to disclose the integrable structure underlying both the gauge theory and its string dual (for pedagogical reviews see for instance 
\cite{Alday:2008yw,Schabinger:2011kb,Adamo:2011pv}). Similar properties have also emerged \cite{BLM}-\cite{BGLP2} in the three dimensional superconformal cousin of ${\cal N}=4$ SYM, the so-called  ABJ(M) theory \cite{ABJM,ABJ}. 

Supersymmetric Wilson loops in $U(N) \times U(M)$ ABJ(M) theory can be constructed \cite{Drukker} as the holonomy of a generalized gauge connection. It naturally includes a non--trivial coupling to the scalars of the form ${\cal M}_J^{\ \  I}(\tau) C_I \bar{C}^J$, governed by a matrix which is locally defined along the path. When ${\cal M}$ is constant,
${\cal M} = {\rm diag}(1,1,-1,-1)$ and the path is chosen to be a maximal circle on $S^2$,  we obtain the well studied $1/6-$BPS Wilson loop $W^{1/6}$ \cite{Drukker, Chen, Rey}. Adding local couplings to the fermions allows to generalize the Wilson operator to the holonomy of a superconnection of the $U(N|M)$ supergroup, leading to an enhanced $1/2-$BPS operator $W^{1/2}$ \cite{DT} (see also  \cite{Lee:2010hk} for an alternative derivation and \cite{Berenstein:2008dc} for previous attempts).    

Perturbative results for $1/6-$BPS Wilson loops \cite{Chen,Rey, BGLP, BGLP3} on the maximal circle have been proved to match the exact prediction obtained by using localization techniques \cite{Kapustin}. At variance with ${\cal N}=4$ SYM \cite{Pestun:2007rz}, the corresponding matrix model is no longer gaussian due to non--trivial contributions from the vector and the matter multiplets. In \cite{MarinoPutrov,Drukker:2010nc} the exact quantum value of this Wilson loop has been obtained by evaluating the matrix model through topological string theory techniques. These results have been further generalized \cite{Klemm:2012ii} using a powerful Fermi gas approach \cite{Marino:2011eh}. The strong coupling limit of the exact expressions matches the predictions from the AdS dual description. 

The fermionic $1/2$--BPS Wilson loop has been proved to be cohomologically equivalent to a linear combination of $1/6$--BPS Wilson loops, since their difference is expressible as an exact ${\cal Q}$--variation, where ${\cal Q}$ is the SUSY charge used in localizing the functional integral of the $1/6-$BPS operator \cite{DT}. Therefore, its vev localizes to the same matrix model and a prediction for its exact value can be easily obtained from the $1/6-$BPS vev\footnote{Actually, as remarked in \cite{MarinoPutrov,Drukker:2010nc}, the relevant linear combination is easier to calculate.}. Perturbative results \cite{BGLP, BGLP3, GMPS} not only agree with this prediction but also confirm the correct identification of the framing factor \cite{Witten} arising from the matrix model calculation \cite{Kapustin}. It is interesting to note that in the $1/2-$BPS case the appearance at perturbative level of non--trivial contributions from the fermionic sector is instrumental to recover the correct framing factors.
   
A more general class of fermionic Wilson loops $W_F[\Gamma]$ living on arbitrary contour $\Gamma$ on $S^2$ has been introduced in \cite{Cardinali}. They are characterized by a non--constant ${\cal M}(\tau)$ and depend on an internal angular parameter $\alpha$. They should be considered the most direct three--dimensional analogue of the DGRT Wilson loop in four dimensions \cite{Drukker:2007dw,Drukker:2007qr,Drukker:2007yx}. Particular representatives within this family $W_F(\alpha,\theta_0)$ have contour on a latitude at an angle $\theta_0$. They generalize the corresponding four--dimensional operators constructed in \cite{Drukker:2006ga} and are in general $1/6-$BPS\footnote{Recently, a bosonic $\th_0$--latitude Wilson loop has been also considered \cite{Marmiroli}, which seems to share quantum features with the latitude operator in four dimensions. In particular, quantum results seem to be related to the ones for $1/6-$BPS Wilson loop simply by a shift  $\l \to \l \cos^2{\th_0}$ in the coupling constant.}. For $\alpha = \tfrac{\pi}{4}$ we are back to the $1/2-$BPS operator of \cite{DT}, whereas for $\alpha = 0$ a new class of three--dimensional Zarembo--like Wilson loops \cite{zarembo} are obtained. 

As in the $\a =\tfrac{\pi}{4}$ case, the fermionic Wilson loop has a bosonic counterpart $W_B(\alpha,\theta_0)$ where the fermionic couplings are set to zero, while the bosonic ones correspond to a latitude coupling encoded into a block--diagonal, path--dependent matrix $\widehat{\cal M}$. For latitude loops these are in general $1/12-$BPS operators, whereas on the equator and for $\a =\tfrac{\pi}{4}$ they reproduce the bosonic $1/6-$BPS Wilson loop of \cite{Drukker}. 

In this paper we begin a detailed investigation at quantum level of these two classes of Wilson loops for which no results are yet available in the literature. 

First of all, at classical level we discuss the cohomological equivalence between the fermionic latitude Wilson loop $W_F$ and the bosonic ones $W_B, \hat{W}_B$ associated to the two gauge groups, and in both cases we determine the number of preserved supersymmetries.  
Then, for both operators defined on a generic $\theta_0$--latitude circle in $S^2$ we perform a two--loop evaluation of their vacuum expectation value. The results (see eqs. \eqref{bosonicW}, \eqref{eq:Wffinal}) exhibit a number of interesting features that we now summarize. 

\begin{itemize}
\item
First of all, although these operators depend on two different parameters, the geometrical latitude $\th_0$ on $S^2$ and the internal angle $\a$, they can be defined in terms of a single combination of the two
\beq
\nu \equiv \sin{2\alpha} \cos{\theta_0}
\eeq
Their expectation value is therefore a function of the coupling and the parameter $\nu$.
Hence we shall refer to the fermionic and bosonic latitude operators as $W_F(\nu)$ and $W_B(\nu)$, respectively.

Setting $\nu = 1$ we expect to enhance the supersymmetry and recover the previously known BPS configurations. For this particular value, in fact, the result for the fermionic Wilson loop collapses to the one of the $1/2-$BPS \cite{DT}, while the result for the new bosonic Wilson loop reduces to the two--loop contribution to the $1/6-$BPS \cite{Drukker}.  

Instead, for $\nu \to 0$ (Zarembo--like limit or, equivalently, path shrinking to the north pole) they both reduce to the two--loop contribution to an operator in pure $U(|N-M|)$ Chern--Simons theory. This is quite in contrast with the expectation. In fact, in analogy with what happens in ${\cal N}=4$ SYM, one would expect the scalars to decouple, so leading to a pure $U(N)$ (or $U(M)$) Chern--Simons vev. Instead, in the present case a residual effect of matter loops survives, which changes the nature of the theory. 

\item
For generic $\nu$ we find an interesting relation between the perturbative results of the two Wilson loops, which encodes quantum corrections to the classical cohomological equivalence. This generalizes the well--known relation linking the $1/2-$BPS and the $1/6-$BPS vev's when computed perturbatively, at framing zero \cite{DT,Drukker:2010nc,BGLP, BGLP3, GMPS}. 
 
In the undeformed case this relation becomes even simpler when the vev's are given at framing--one, as obtained in the matrix model approach \cite{DT,Drukker:2010nc}. Inspired by this observation and motivated by the search for a putative ``framed'' computation compatible with the cohomological equivalence,  we are led to conjecture that the following identity
\beq
\label{frame}
\langle W_F(\nu)\rangle_\nu = \frac{N\, e^{-\frac{i\pi \nu}{2}}\, \langle W_B(\nu)\rangle_\nu - M\, e^{\frac{i\pi \nu} {2}}\, \langle\hat W_B(\nu)\rangle_\nu}{N\, e^{-\frac{i\pi \nu}{2}}   - M\, e^{\frac{i\pi \nu} {2}}}
\eeq
should hold for ``framing--$\nu$'' quantities\footnote{This is the meaning of the subscript $\nu$.} properly defined in terms of our framing--zero perturbative expectation values. They differ by a $\nu$--dependent phase, according to a prescription that generalizes that for the $\nu=1$ case (see eq. \eqref{bosonicWfram}). 

Relation \eqref{frame} suggests the existence of a matrix model that should arise from a suitable localization of the functional integral\footnote{Localization usually reduces the path-integral to a sum over discrete or continuous constant field configurations, but, in general, it could also lead to a lower dimensional field theory \cite{Pestun:2009nn}.}, and that would provide Wilson loops vev at non--integer framing $\nu$. Searching for this matrix model is certainly challenging. 

\item 
In conformal field theories Wilson loops allow for computing the energy radiated by a moving quark in the low energy limit (Bremsstrahlung function $B(\l)$) \cite{Correa:2012at,Fiol:2012sg} and the contribution to the entanglement entropy due to a heavy quark sitting inside a finite region \cite{Lewkowycz:2013laa}. 
The Bremsstrahlung function also governs the small angle expansion of the cusp anomalous dimension $\G_{cusp}(\varphi, \th) \simeq B(\l) (\th^2 - \varphi^2)$ for a generalized cusp. The parameter $\varphi$ represents the geometric angle between the Wilson lines, whereas $\theta$ accounts for the change in the orientation of the couplings to the scalars between the two rays. In ${\cal N}=4$ SYM theory an exact prescription to extract this non--BPS observable from BPS loops has been given in the seminal paper \cite{Correa:2012at}. The original proposal was further elaborated and substantiated in \cite{Correa:2012nk, Correa:2012hh}. 

Recently, for the ABJM theory a general formula has been proposed \cite{Lewkowycz:2013laa}, which gives the Bremsstrahlung function $B_{1/6}$ for $1/6-$BPS quark configurations as the derivative of a bosonic Wilson loop on a squashed sphere with respect to the squashing parameter $b$. An equivalent expression in terms of the $n$--derivative of a Wilson loop winding $n$ times the great circle (with the dictionary $b=\sqrt{n}$) has also been provided, which is amenable of explicit computations. In \cite{Lewkowycz:2013laa} a proposal for extending the general prescription to the $1/2-$BPS case is also discussed, but the authors leave a number of open questions to be clarified. 

In this paper, we further elaborate on these questions by investigating the possibility of computing Bremsstrahlung functions in terms of our latitude Wilson loops. First of all, supported by the perturbative results we propose 
\beq
\label{prescription}
B_{1/2}(\lambda) = \frac{1}{4\pi^2}\, \partial_\nu\,  \log\, \langle W_F(\nu)\rangle_0\, \bigg|_{\nu=1}
\eeq
as the right prescription for determining the Bremsstrahlung function for the $1/2-$BPS cusp in ABJM, in terms of the fermionic latitude Wilson loop. 

When this equation is applied to our two--loop result at framing zero $\langle W_F(\nu)\rangle_0$, in the planar limit ($\l \equiv N/k$), it agrees with $B_{1/2}(\l)$ as obtained directly from the perturbative computation of the 1/2--BPS generalized cusp \cite{Griguolo}. It is important to stress that already at this order the matching is non--trivial. In fact, when specialized to the ABJM case ($M=N$) our result $\langle W_F(\nu)\rangle_0$ surprisingly looses the $\nu$ dependence in the $\l^2$ coefficient, so leading to a Bremsstrahlung function which at this order is odd in $\l$. On the other hand, this is exactly what we obtain if we compute $B_{1/2}$ directly from the result of \cite{Griguolo} for the $1/2-$BPS cusp.

\item
It is interesting to observe that exploiting the cohomological equivalence \eqref{frame} the previous prescription can be rephrased in terms of bosonic Wilson loops $W_B$ as 
\beq
\label{prescription2}
B_{1/2}(\lambda) = \frac{1}{4\pi^2} \left[\partial_\nu\, \log\left(\langle W_B(\nu)\rangle_\nu+\langle \hat W_B(\nu)\rangle_\nu\right)\bigg|_{\nu=1}\!\!-\frac{i\pi}{2}\frac{\langle W_B(1)\rangle_1-\langle \hat W_B(1)\rangle_1}{\langle W_B(1)\rangle_1+\langle \hat W_B(1)\rangle_1}\right]
\eeq
In particular, since $W_B(1)=W^{1/6}$, the second term can be easily computed from the well--known results of localization for the $1/6-$BPS bosonic Wilson loop on the maximal circle \cite{Kapustin,MarinoPutrov,Drukker:2010nc,Klemm:2012ii} and allows to make an interesting  prediction for $B_{1/2}(\l)$ at three loops 
\beq
B_{1/2}(\lambda) = \frac{\l}{8} - \frac{\pi^2}{48} \l^3 + {\cal O}\left(\l^5\right) 
\eeq
It would be interesting to check this formula against a direct three--loop evaluation of the $1/2-$BPS cusp anomalous dimension.

\item
Inspired by the recipe recently given in \cite{Lewkowycz:2013laa} for computing $B(\l)$ and the similarity between our parameter $\nu$ and the squashing parameter $b$ used there, we are led to conjecture that our prescription \eqref{prescription2} could be rewritten in terms of multiply $n$--wound Wilson loops $W^{1/6}_n$ whose vev is known exactly from localization. Formally setting $n = n(\nu)$ with $n(1)=1$ and taking into account that $\langle W^{1/6}_1\rangle = \langle W_B(1)\rangle_1$, we  rewrite 
\beq
\label{nformula}
B_{1/2}(\l) = \frac{1}{4\pi^2} \left[ \partial_n\, 
\log\left(\langle W^{1/6}_n\rangle +\langle \hat W^{1/6}_n\rangle \right)  \frac{\pa n}{\pa \nu}\Bigg|_{n=1}
-\frac{i\pi}{2}\, \frac{\langle W^{1/6}_1\rangle -\langle \hat W^{1/6}_1\rangle}{\langle W^{1/6}_1\rangle +\langle \hat W^{1/6}_1\rangle}
 \right]
\eeq
We have checked this proposal using the weak and strong coupling expansion of the exact expression for $\langle W^{1/6}_n\rangle$ given in \cite{Klemm:2012ii}. At weak coupling we reproduce exactly the  two--loop result from the fermionic cusp. At strong coupling the leading term coincides with the one of \cite{Forini:2012bb}, while the first subleading term does not\footnote{The mismatch in the subleading term might be due to the nature of the result in \cite{Forini:2012bb}, which does not seem to respect the BPS condition.}. 

An interesting pattern seems to emerge when applying the recipe \eqref{nformula}. Using the expansions of $\langle W^{1/6}_n\rangle$, it turns out that both at weak and strong coupling the functional dependence of the coefficients on $n$ is such that the first term in \eqref{nformula} always vanishes, and the actual expression for $B_{1/2}(\l)$ is totally encoded in the second term. A similar pattern arises also in the proposal of \cite{Lewkowycz:2013laa} for the Bremsstrahlung function in the fermionic case. In particular, this leads to the conclusion that $B_{1/2}(\l)$ should be described by an odd function of $\lambda$, although we agree with the authors of \cite{Lewkowycz:2013laa} that the physical meaning of this result has  still to be fully understood. On the other hand, at least up to two loops, it is supported by the explicit calculation of the cusp \cite{Griguolo}. 

\item
We can try to generalize the recipe \eqref{prescription} to the bosonic case. If we apply the derivative with respect to $\nu$ to the vev $\langle W_B(\nu) \rangle_0$ we expect to reproduce the two--loop result for the Bremsstrahlung function in the $1/6-$BPS quark configurations, as can be read from the weak coupling expansion of the $1/6-$BPS cusp
\cite{Griguolo}. Actually, we find a result that differs from the correct one by a factor $1/2$. We understand this mismatch as coming from the fact that in the $1/6-$BPS case the cusp anomalous dimension does not satisfy the BPS condition $\G_{1/6}(\varphi=\th) = 0$.  
\end{itemize}

\vskip 5pt

The paper is organized as follows. In section 2 we introduce the generalized fermionic Wilson loop of \cite{Cardinali} and discuss its ${\cal Q}$--equivalence with a new kind of bosonic latitude operators. 
In section 3 we present the two--loop evaluation of both Wilson loops, while in section 4 we discuss in detail the relation between the two results in terms of a non--integer framing $\nu$, the calculation of the Bremsstrahlung function and the connection with the recent proposals of \cite{Lewkowycz:2013laa}. Few appendices follow, which contain conventions and details of the perturbative calculation.

\section{Generalized Wilson loops}
\label{sec:generalities}

\subsection{Fermionic latitude}
In \cite{Cardinali}  it was shown  that we can associate  a supersymmetric Wilson loop operator  to any contour  lying on the two dimensional sphere:  $x^\mu x_\mu=1$.  The key idea is to  embed  the original $U(N)\times U(M)$ connection  present in ABJ theories into an {\it effective} $U(N|M)$ superconnection given by (for our conventions on Chern--Simons matter theories see appendix \ref{sec:conventions})
\bea
\label{supermatrix}
\mathcal{L} &=& \begin{pmatrix}
\mathcal{A}
&-i \sqrt{\frac{2\pi}{k}}  |\dot x | \eta_{I}\bar\psi^{I}\\
-i \sqrt{\frac{2\pi}{k}}   |\dot x | \psi_{I}\bar{\eta}^{I} &
\hat{\mathcal{A}}
\end{pmatrix} \ \  \  \mathrm{with}\ \ \  \left\{\begin{matrix} \mathcal{A}\equiv A_{\mu} \dot x^{\mu}-\frac{2 \pi i}{k} |\dot x| {\cal M}_{J}^{\ \ I} C_{I}\bar C^{J}\\
\\
\hat{\mathcal{A}}\equiv\hat  A_{\mu} \dot x^{\mu}-\frac{2 \pi i}{k} |\dot x| {\cal M}_{J}^{\ \ I} \bar C^{J} C_{I}
\end{matrix}\ \right.
\non \\
&\equiv&  \mathcal{L}_B + \mathcal{L}_F
\eea
Here we have called $\mathcal{L}_B = \mathrm{diag}(\mathcal{A},\hat{\mathcal{A}})$, while $\mathcal{L}_F$ is the off--diagonal fermionic matrix. 

The matrix   $ {{\cal M}}_{J}^{\ \ I}$ governing the coupling to the scalar fields can be expressed in terms of the contour $x^\mu$ as follows  
\begin{align}
\label{boscoup}
{\cal M}_{K}^{\ \ J} =
\ell\delta^{J}_{K}-2i \ell s_{K}\bar s^{J}\!\!-2 i  \cos 2\alpha ~\!s_{K}\frac{\dot{x}\cdot \gamma}{|\dot x|} \bar s^{J}~\!\!\!-2 i\! \sin2\alpha~s_{K}\gamma^{\lambda}\bar s^{J}~\epsilon_{\lambda\mu\nu} x^{\mu}\frac{\dot x^{\nu}}{|\dot x|}
\end{align}
while the Grassmann even  spinors  $(\eta_{I}^{\beta},\bar\eta^I_\beta)$ which control the fermionic couplings are given by
\begin{subequations}
\label{fermcoup}
\begin{align}
\eta^{\beta}_{I}=& \,
i\, e^{\frac{i}{2}\ell(\sin2\alpha) \tau}\left[s_{I}(\cos\alpha~\mathds{1}-i \sin\alpha~ (x^{\mu}\gamma_{\mu}))\left(\mathds{1}+\ell\frac{\dot{x}\cdot \gamma}{|\dot x|}\right)\right]^{\beta}
\\
\bar\eta_{\beta}^{I}=& \,
i\, e^{-\frac{i}{2}\ell(\sin2\alpha) \tau}\left[\left(\mathds{1}+\ell\frac{\dot{x}\cdot \gamma}{|\dot x|}\right)\left(\cos\alpha~\mathds{1}+i \sin\alpha~ (x^{\mu}\gamma_{\mu})\right)\bar s^{I}\right]_{\beta}
\end{align}
\end{subequations}
The parameter $\tau$ appearing in the exponent  is the affine parameter of the curve, $\gamma^\lambda$ are the euclidean Dirac matrices in three dimensions, while  $s_I $ and $\bar s^I$  denote two sets of constant bosonic spinors obeying the orthogonality relation
 \beq
 \label{dfg}
 \bar s^{I}_{\beta}\, s_{I}^{\alpha}=\frac{1}{2 i}\, \delta^{\alpha}_{\beta}
 \eeq
The angle $\alpha$  can be freely chosen in the interval $\left[0,\frac{\pi}{2}\right]$. If our space-time were a sphere $S^3$, this quantity would represent the relative position of our $S^2$ inside $S^3$. The constant parameter $\ell$ in \eqref{boscoup} and \eqref{fermcoup} can only take two values, $\pm1$, and  its  choice specifies the eigenvalues of the matrix $\cal M$: $(-1,1,1,1)$  $[\ell =1]$ and $(1,-1,-1,-1) $ $[\ell=-1]$. 

The existence of  superconformal transformations  preserving the Wilson loops defined in \eqref{supermatrix}  is 
discussed  in detail in \cite{Cardinali}. There,  it was shown  that some  of the supercharges, when acting on the holonomy defined by $\mathcal{L}$,
are  realised  as  $U(N|M)$ supergauge transformations, namely 
\beq
\mathcal{W}[\Gamma]=\mathrm{P}\,\exp\left(-i\oint_\Gamma\mathcal{L}(\tau)d\tau\right)\ \ \  \xmapsto{\  \  \  \mbox{ \large $\mathcal{Q}$}\   \  \ }\ \ \ \ \mathcal{W}^\prime[\Gamma] = U(2\pi)\, \mathcal{W}[\Gamma]\, U^{-1}(0)
\eeq
However, the supertrace\footnote{Recall the we are dealing with supermatrices  and only supertraces are invariant under cyclic permutations of their arguments.} of $\mathcal{W}$ does not yield  a supersymmetric operator since the supergauge transformation $U$ is not periodic but it obeys the twisted boundary condition  $U(2\pi)=\mathcal{T}^{-1} U(0) \mathcal{T}$ with
\beq
\mathcal{T}=\begin{pmatrix} e^{-\frac{i\ell }{4}(\sin2\alpha) L}  \mathds{1}_N& 0\\ 0 & e^{\frac{i\ell}{4}(\sin2\alpha) L}  \mathds{1}_M\end{pmatrix}
\eeq
where $L$ stands for  the perimeter of the contour. The failure of periodicity of $U$ stems from the two phases present in the fermionic couplings \eqref{fermcoup}.  Therefore, a gauge invariant operator is defined by explicitly inserting the matrix $\mathcal{T}$ in the supertrace
\beq
\label{eq:fermionic}
W_F[\Gamma]=\frac{\mathrm{STr}(\mathcal{W}[\Gamma] \mathcal{T})}{\mathrm{STr}(\mathcal{T})}  \,\equiv \, {\cal R} \;  \mathrm{STr}(\mathcal{W}[\Gamma] \mathcal{T})
\eeq

This operator generically preserves two superconformal supercharges leading to a $1/12-$BPS Wilson loop. 
If we choose $x^\mu$ to be the equatorial circle and set $\alpha=\frac{\pi}{4}$ we recover the $1/2-$BPS circle  introduced in \cite{DT} as one of the elements of this larger family. 

Apart from the case of the equatorial  circle \cite{DT} nothing is known  about the quantum  properties of this class of Wilson loops.  Here we start their investigation by considering the simplest (but non--trivial) generalization of the equator,  namely the {\it latitude} on $S^2$
\beq
\label{latitude}
x^\mu = (\sin\theta_0, \cos\theta_0 \cos \tau, \cos\theta_0 \sin\tau)  \  \  \ \mbox{with}  \  \  \    -\frac{\pi}{2}\le\theta_0\le \frac{\pi}{2}
\eeq 
In this case  the  general form \eqref{fermcoup} and \eqref{boscoup}  of the couplings is greatly  simplified. With a suitable  choice\footnote{\label{foot1}For instance,  we can select 
$s_I^\alpha= u_I \rho^\alpha+v_I \bar\rho^\alpha$ and  $\bar s^I_\alpha= -i (s_I^\alpha )^\dagger,
$
where $u_I=(e^{i\delta},0,0,0)$ and $v_I=( 0,1,0,0)$.  The two spinors $\rho^\alpha$ and $\bar \rho^\alpha$ can be taken to be eigenstates of $\sigma_3$: $\rho^{\alpha}=\frac{i }{\sqrt{2}}\left( e^{i \gamma},0\right)$  and $\bar\rho^{\alpha}=\frac{1 }{\sqrt{2}}\left( 0,-e^{-i \gamma}\right)$. The  expression \eqref{eq:matrixfermionic} is then  obtained by setting  $\tan\delta=\frac{\sin 2\theta_0 }{\cot ^2\alpha -\cos 2 \theta_0 }$ and $\tan\gamma=\frac{\sin \theta_0 }{\cos \theta_0 -\cot \alpha }$. Different choices for the vectors $u_{I}$ and $v_{I}$ and the spinors $\rho_{\alpha}$ and $\bar\rho^{\alpha}$ would lead to equivalent forms of the couplings. In fact, they only differ by global Lorentz and $R-$symmetry rotations. } of the constant spinors $s_{I}$ and $\bar s^{I}$ we can always realize the following representation
\bea
\label{eq:matrixfermionic}
\mbox{\small $\!{\cal M}_{I}^{ \  J}\!=\!\!\left(\!\!
\begin{array}{cccc}
 - \nu  & e^{-i \tau } 
   \sqrt{1-\nu ^2} & 0 & 0 \\
e^{i \tau }  \sqrt{1-\nu ^2}
   & \nu  & 0 & 0 \\
 0 & 0 & 1 & 0 \\
 0 & 0 & 0 & 1 \\
\end{array}\!
\right)$} \quad &,& \quad  \mbox{\small $\begin{array}{l}\eta_I^\alpha \equiv n_I \eta^\a = \frac{e^{\frac{i\nu \tau}{2}}}{\sqrt{2}}\left(\!\!\!\begin{array}{c}\!\sqrt{1+\nu}\\ -\sqrt{1-\nu} e^{i\tau}\\0\\0 \!\end{array}\!\!\right)_{\!I} \! \! \!\!(1, -i e^{-i \tau})^\alpha
\end{array}$}\!\!
\non \\
&~& \quad \bar\eta_\alpha^I \equiv \bar{n}^I \bar{\eta}_\a = i (\eta^{\alpha}_{I})^{\dagger}
\eea
where  $\nu \equiv \sin 2\alpha\cos\theta_0$ and we have set $\ell=1$ to stick to the conventions of \cite{DT}. Note that  also the matrix $\mathcal{T}$  and  consequently the normalization ${\cal R} = 1/\mathrm{Str}(\mathcal{T})$ in \eqref{eq:fermionic} depend only on  the parameter  $\nu$
\beq
\label{eq:T}
\mathcal{T}=\begin{pmatrix} e^{-\frac{i\pi\nu }{2}}  \mathds{1}_N & 0\\ 0 & e^{\frac{i\nu\pi}{2}}  \mathds{1}_M\end{pmatrix} \  \  \  \  \mathrm{and} \  \  \  \ {\cal R} =  \frac{1}{N e^{-\frac{i\pi\nu }{2}} -M e^{\frac{i\pi\nu }{2}} }
\eeq
In the limit $\nu\to 1$ we recover all the known results of the $1/2-$BPS circle.   

To determine the number of supersymmetries  preserved by  a generic latitude  we have  to  solve the  general  set of  BPS conditions given in  \cite{Cardinali}
 \begin{subequations}
\begin{alignat}{4}
\label{cond1}
&\mathrm{(A):}\ \ \ \ \epsilon_{IJKL} (\eta\bar{\Theta}^{IJ})\bar n^{K}=0\ \ \ \ \ \ &\mathrm{and}\ \ \  \ \ \ \ \  &\mathrm{(B):}\ \  \ \ \
n_{I}(\bar\eta\bar\Theta^{IJ})=0
\\
&\mathrm{(A):}\ \ \ \ \bar{\Theta}^{IJ}\partial_{\tau} {\bar{\eta}}^K\epsilon_{IJKL}=0
\ \ \ \ \ \ &\mathrm{and}\ \ \  \ \ \ \ \  
&\mathrm{(B):}\ \  \ \ \
\bar\Theta^{IJ}\partial_{\tau}{\eta}_{I}=0
\end{alignat}
\end{subequations}
where in flat space the constant  spinors $\bar\Theta^{IJ}$ are defined in terms of the supersymmetry ($ \bar\theta^{IJ}$) and superconformal ($\bar\epsilon^{IJ}$) parameters as  
$\bar\Theta^{IJ}=\bar\theta^{IJ}-(x\cdot\gamma)\bar\epsilon^{IJ}$.

Introducing the two independent combinations
$\bar \chi^{IJ}=\bar{\theta}^{IJ}+i e^{i\theta_0}\gamma_{1}\bar\epsilon^{IJ} $  {and}  $\bar\kappa^{IJ}=\bar{\theta}^{IJ}-i e^{-i \theta_0 }\gamma_{1}\bar\epsilon^{IJ}$, after a long and tedious spinor algebra the above constraints can be reduced to  the following two sets of equations
\begin{subequations}
\label{eqforchi}
\begin{align}
\!\!\!
\mbox{\sc Eqs for $\bar\chi^{IJ}$:}\  \  \  \  
&s_{I}
 (\cos\alpha- e^{i\theta_0}\sin\alpha\gamma_{1} )\gamma_{\mu}\bar \chi^{IJ}=0\\
&\label{Sa}\epsilon_{IJKL}\bar s^K(\cos\alpha +\sin\alpha e^{-i\theta_0} \gamma_1)(\nu-\gamma_{1}) \gamma_i\bar \chi^{IJ}=0\ \ \  (i=2,3)\\
&\epsilon_{IJKL}\bar s^K(\cos\alpha +\sin\alpha e^{-i\theta_0} \gamma_1) (\nu-\gamma_1)\bar \chi^{IJ}=0
\end{align}
\end{subequations}
\vskip -.5truecm
\begin{subequations}
\label{eqforkappa}
\begin{align}
\!\!\!\!\!\!\!\!\!\!\!\!\!\!\!
\mbox{\sc Eqs for $\bar\kappa^{IJ}$:}\  \  \  \  
&\epsilon_{IJKL} 
\bar s^{K} (\cos\alpha+ e^{-i\theta_0}\sin\alpha\gamma_{1} )\gamma_{\mu} \bar\kappa^{IJ}=0\\
&s_{I} (\cos\alpha- \sin\alpha e^{i\theta_0}\gamma_{1})(\nu+\gamma_{1}) \gamma_i  \bar\kappa^{IJ}=0 \ \  \  \  \  (i=2,3)\\
 &s_{I} (\cos\alpha- \sin\alpha e^{i \theta_0}\gamma_{1})(\nu+\gamma_{1}) \bar\kappa^{IJ}=0
\end{align}
\end{subequations}
The linear systems of equations  \eqref{eqforchi} and \eqref{eqforkappa}  can be  solved  by using the expansion of 
$s_I$ and $\bar s^I$  suggested  in footnote \ref{foot1}.   The general  solution is parametrized by four constants $\omega_i$  ($i=1,..,4$) and 
the only non vanishing components of  $\bar\Theta^{IJ}$ (up to the obvious antisymmetry $\bar\Theta^{IJ}=-\bar\Theta^{JI}$) are  given by
\beq
\label{chargferm}
\begin{split}
\!\!\!\bar\theta^{13}_1=&e^{-\frac{i \theta_0 }{2}} \sqrt{1-\nu } ~\omega _1+e^{\frac{i \theta_0
   }{2}} \sqrt{1+\nu} ~\omega _2  \ \ \  \ \  \  \  \  \ ~\bar\theta^{14}_1=e^{-\frac{i \theta_0 }{2}} \sqrt{1-\nu } ~\omega _3+e^{\frac{i \theta_0
   }{2}} \sqrt{1+\nu} ~\omega _4 \\
\!\!\!\bar\theta^{23}_2=& -i e^{-\frac{i \theta_0 }{2}} \sqrt{1+\nu} ~\omega _1-i e^{\frac{i
   \theta_0 }{2}} \sqrt{1-\nu } ~\omega _2 ~   \  \  \ \bar\theta^{24}_2= -i e^{-\frac{i \theta_0 }{2}} \sqrt{1+\nu} ~\omega _3-i e^{\frac{i
   \theta_0 }{2}} \sqrt{1-\nu } ~\omega _4 \\
\!\!\!{\bar\epsilon}^{13}_1=&i e^{\frac{i \theta_0 }{2}} \sqrt{1-\nu } ~\omega _1-i e^{-\frac{i \theta_0
   }{2}} \sqrt{1+\nu} ~\omega _2 \ \ \ \ \  \  \  \  {\bar\epsilon}^{14}_1=i e^{\frac{i \theta_0 }{2}} \sqrt{1-\nu } ~\omega _3-i e^{-\frac{i \theta_0
   }{2}} \sqrt{1+\nu} ~\omega _4 \\
\!\!\!{\bar\epsilon}^{23}_2=& e^{-\frac{i \theta_0 }{2}} \sqrt{1-\nu } ~\omega _2-e^{\frac{i \theta_0
   }{2}} \sqrt{1+\nu} ~\omega _1 \ \ \  \ \ \ \  \  \  \ {\bar\epsilon}^{24}_2=e^{-\frac{i \theta_0 }{2}} \sqrt{1-\nu } ~\omega _4-e^{\frac{i \theta_0
   }{2}} \sqrt{1+\nu} ~\omega _3 
\end{split}
\eeq
Thus the fermionic latitude defined by $\eqref{eq:matrixfermionic}$  is  ${1/6}$-BPS. The usual ${1/2}$-BPS circle is recovered by
setting $\alpha=\frac{\pi}{4}$ and $\theta_0=0$ ({\it i.e.}  $\nu=1$). In fact, for this choice of the parameters the last two equations in \eqref{eqforchi} and \eqref{eqforkappa} are identically satisfied and the supersymmetry is enhanced from $1/6$  to $1/2$.

\vskip 10pt 
\subsection{Bosonic latitude}
Given the $1/6-$BPS fermionic  latitude  \eqref{eq:matrixfermionic} we can introduce its  bosonic 
version with local $SU(2)\times  SU(2)$ symmetry. This is defined as the holonomy of  the $U(N)$ connection
\beq
\label{eq:matrixbosonic}
 \mathcal{L}_b\equiv A_{\mu} \dot x^{\mu}-\frac{2 \pi i}{k} |\dot x| \widehat{{\cal M}}_{J}^{\ \ I} C_{I}\bar C^{J}
\  \  \    \mathrm{with}\ \ \ 
 \mbox{\small $\!\widehat{{\cal M}}_{J}^{\ I}=\left(\!\!
\begin{array}{cccc}
 - \nu  & e^{-i \tau } 
   \sqrt{1-\nu ^2} & 0 & 0 \\
e^{i \tau }  \sqrt{1-\nu ^2}
   & \nu  & 0 & 0 \\
 0 & 0 & -1 & 0 \\
 0 & 0 & 0 & 1 \\
\end{array}\!
\right)$}
\eeq
Alternatively we can use  its  $U(M)$ analogue, $\hat{\mathcal{L}}_b\equiv\hat  A_{\mu} \dot x^{\mu}-\frac{2 \pi i}{k} |\dot x| \widehat{{\cal M}}_{J}^{\ \ I} \bar C^{J} C_I$, with the 
same matrix   $\widehat{{\cal M}}_{J}^{\ I}$. 

The bosonic loop operator defined by \eqref{eq:matrixbosonic} is again supersymmetric.  In this case the  BPS condition  $\delta_{\bar\Theta}  \mathcal{L}_b=0$ can be shown to be equivalent to the following set of constraints for the supercharge
\begin{subequations}
\begin{align}
\label{condB1}
&\!\!\!(\eta\bar\Theta^{IJ})+\widehat{{\cal M}}_K^{\ \  I}(\eta\bar\Theta^{KJ})=0 \ \  \  \  \  \ \ \ \ \   \  \  \  \  \ \ \  (\bar\eta\bar\Theta^{IJ})-\widehat{{\cal M}}_K^{\ \  I}(\bar\eta\bar\Theta^{KJ})=0\\
\label{condB2}
&\!\!\!\epsilon_{IJKR} (\eta \bar\Theta^{IJ}) + \widehat{{\cal M}}_R^{\  \ S}\epsilon_{IJSK}
(\eta \bar\Theta^{IJ})=0
 \  \  \  \epsilon_{IJKR}
 (\bar\eta \bar\Theta^{IJ}) - \widehat{{\cal M}}_R^{\  \ S }\epsilon_{IJSK}
(\bar\eta \bar\Theta^{IJ})=0
\end{align}
\end{subequations}
where the spinors $\eta$ and $\bar\eta$ are the two  eigenstates of the matrix $(\dot x\cdot\gamma)$:  $
 (\dot{x}^{\mu}\gamma_{\mu})\bar\eta= |\dot x| \bar\eta,\ \ \mathrm{and} \ \ 
 (\dot{x}^{\mu}\gamma_{\mu})\eta=- |\dot x| \eta.$ They provide a natural basis  for the spinors. It is easy to realize that the second set of conditions is automatically satisfied once    eqs. \eqref{condB1} hold\footnote{This can be shown by dualizing the two equations in \eqref{condB2}. Since the trace of $\widehat{{\cal M}}$ is zero, we obtain an expression which identically vanishes when eqs.\ \eqref{condB1} are satisfied.}. The remaining two equations can be explicitly solved and in terms of two spinorial parameters the general solution reads
\beq
\label{chargbos} 
\begin{array}{ll}
\bar\theta^{13}_1=e^{\frac{i \theta_0 }{2}} \sqrt{1+\nu} ~\zeta _1 ~~~~&
\bar\theta^{14}_1= e^{-\frac{i \theta_0 }{2}} \sqrt{1-\nu }~\zeta _2  \\ 
\bar\theta^{23}_2=-i e^{\frac{i \theta_0 }{2}} \sqrt{1-\nu }~\zeta _1  ~~~~&
\bar\theta^{24}_2=-i e^{-\frac{i \theta_0 }{2}} \sqrt{1+\nu}~\zeta _2 \\  
\bar\epsilon^{13}_1=-i e^{-\frac{i \theta_0 }{2}} \sqrt{1+\nu} ~ \zeta _1~~~~&
\bar\epsilon^{14}_1=i e^{\frac{i \theta_0 }{2}} \sqrt{1-\nu }~\zeta _2 \\ 
\bar\epsilon^{23}_2= e^{-\frac{i \theta_0 }{2}} \sqrt{1-\nu }~\zeta _1 ~~~~&
\bar\epsilon^{24}_2= -e^{\frac{i \theta_0 }{2}}\ \sqrt{1+\nu}~\zeta _2
\end{array}
\eeq
Therefore the loop operators
\begin{align}\label{eq:bosonic}
W_B(\nu) \equiv \frac{1}{N}\, \Tr\, \exp\left( \oint_{\G_{\nu}} d\tau\, {\cal L}_b(\tau) \right)
\quad ,  \quad 
\hat{W}_B(\nu) \equiv \frac{1}{M}\, \Tr\, \exp\left( \oint_{\G_{\nu}} d\tau\, \hat{\cal L}_b(\tau) \right)
\end{align}
are both $1/12$-BPS.
We note that the solution \eqref{chargbos}  spans a subset of the supercharges of the fermionic latitude obtained by setting $\omega_1=\omega_4=0$ in  \eqref{chargferm}.

\subsection{The cohomological equivalence}
Consider now the following combination of  the Poincar\'e  $Q^{IJ,\alpha}$ and conformal $S^{IJ,\alpha}$ supercharges defined in \eqref{chargbos}
\beq
\label{supercharge}
\begin{split}
\mathcal{Q}=&-\sqrt{\frac{1+\nu
    }{2}}\left(e^{\frac{i \theta_0 }{2}}
  Q^{13,1}-i e^{-\frac{i \theta_0 }{2}}
  S^{13,1}+e^{-\frac{i \theta_0 }{2}}
 Q^{24,2}-i e^{\frac{i \theta_0 }{2}}
S^{24,2}\right)+\\
   &+i \sqrt{\frac{1-\nu  }{2} }\left(e^{\frac{i \theta_0 }{2}}
Q^{23,2}+i e^{-\frac{i \theta_0 }{2}}
S^{23,2} -e^{-\frac{i \theta_0
   }{2}}Q^{14,1}-i e^{\frac{i \theta_0 }{2}}
 S^{14,1}\right).
 \end{split}
\eeq
This supercharge can be used to relate the fermionic latitude Wilson loop \eqref{supermatrix} with the choice \eqref{eq:matrixfermionic} to the bosonic ones \eqref{eq:bosonic} with the choice \eqref{eq:matrixbosonic}. 
To begin with, we observe that the fermionic part of the superconnection \eqref{supermatrix} is $\mathcal{Q}-$exact, namely
\beq
\label{QL1}
\mathcal{L}_F = \mathcal{Q}\Lambda
\  \   \  \ \mathrm{where}\  \  \  \
\Lambda=i \sqrt{\frac{\pi }{2\kappa}} \left(
\begin{array}{cc}
 0 &  e^{\frac{i \nu  \tau }{2}-\frac{i \theta_0
   }{2}} C_3  \\
e^{\frac{i \theta_0 }{2}-\frac{i \nu  \tau
   }{2}}  \bar{C}^3 & 0 \\
\end{array}
\right)
\eeq
From the above relation and the fact that the fermionic loop is invariant under this supersymmetry transformations we can also show
that 
\beq
\label{QL2}
\mathcal{Q}(\mathcal{L}_F)=8 i D_\tau (|\dot x| \Lambda) \ \  \mathrm{and}\ \ \ \   8 i  |\dot x|\Lambda\Lambda=\mathcal{L}_{B}-\begin{pmatrix}\mathcal{L}_{b} & 0\\  0 & \hat{\mathcal{L}}_{b}\end{pmatrix}
\eeq
With the help of the building blocks \eqref{QL1} and \eqref{QL2} we can straightforwardly  repeat the same path discussed  in  \cite{DT}  for the case of the circle and show that the fermionic latitude is cohomologically equivalent to  the combination
\beq
\label{QL3}
W_B^+(\nu) = 
{\cal R} \left[ N e^{-\frac{\pi i \nu}{2}}W_B(\nu) -M  e^{\frac{\pi i \nu}{2}}\hat W_B(\nu) \right]
\eeq
where ${\cal R}$ is the normalization factor defined in \eqref{eq:T}.
Therefore, classically we can write
\beq 
\label{QL10}
W_F(\nu) - W_B^+(\nu) = \mathcal{Q}V
\eeq
where $V$ is a function of $\L$ and the (super)connections. In section \ref{sec:discussion} we will discuss how this relation is realized perturbatively at quantum level.

An important remark is in order. The operator \eqref{supercharge} which realizes the cohomological equivalence \eqref{QL10} for generic values of the parameter $\nu$ is not simply a $(\alpha,\theta_0)-$dependent deformation of the one used in \cite{DT} to prove the $\mathcal{Q}-$equivalence between the fermionic $1/2-$BPS circle and its bosonic $1/6-$BPS counterpart. In fact, in that case the chosen supercharge is chiral, while  in our case it is never possible to write the $\mathcal{Q}-$equivalence in terms of purely chiral supercharges. Only for $\nu \to 1$ it turns out to be possible to trade the right hand side of \eqref{QL10} for $\mathcal{Q}_c \tilde{\L}$ with $\mathcal{Q}_c$ chiral, thanks to the enhancement of supersymmetry gained in this limit. Therefore, it follows that we cannot use the framework introduced in \cite{Kapustin} for localizing the $1/2-$BPS  circle to evaluate the latitude operator away from $\nu=1$.

\section{Perturbative evaluation}
\label{sec:perturbative}

At weak coupling we can evaluate the vacuum expectation values of Wilson loops perturbatively by Taylor expanding the exponential of the (super)connection and Wick contracting the fields. Below we shall  compute the vev of the generalized fermionic  Wilson loop \eqref{eq:fermionic} and of the bosonic one \eqref{eq:bosonic} up to two loops. This requires  expanding the path-ordered exponential up to the fourth order. In this process, for  the general case of the superconnection \eqref{supermatrix}, we generate purely bosonic contributions from the diagonal part of the $U(N|M)$ supermatrix \eqref{supermatrix}, purely fermionic ones from the off-diagonal blocks and mixed terms from the combination of the two. The  vev of  the bosonic Wilson loop \eqref{eq:bosonic} can be obtained by turning off fermions in the previous analysis. Therefore, we focus on the calculation of the  fermionic Wilson loop \eqref{eq:fermionic} from which we can read both results. 

We consider the generalized BPS Wilson loops \eqref{eq:fermionic} on a latitude circle on $S^2$ parametrized as in (\ref{latitude}).
Given the particular structures of the field propagators (see appendix \ref{sec:conventions}), short distance divergences may arise in loop integrals and in integrations along the contour.  
We regularize them by using the DRED scheme (dimensional regularization with dimensional reduction) \cite{Siegel}, which preserves gauge invariance and supersymmetry \cite{Chen:1992ee}.  

According to the DRED prescription we assign Feynman rules in three dimensions and perform all tensor manipulations strictly in three dimensions before analytically continuing loop integrals to $D= 3-2\epsilon$. Specific rules are then required for contracting three--dimensional objects coming from Feynman rules with $D$--dimensional tensors arising from tensor integrals. These rules \cite{Siegel:1980qs} easily follow from requiring $\e >0$ 
\beq
\label{DRED}
\eta^{\mu\nu} \eta_{\mu\nu} =3 \qquad \quad 
\hat{\eta}^{\mu\nu} \hat{\eta}_{\mu\nu} = 3-2\e \qquad \quad 
\eta^{\mu\nu} \hat{\eta}_{\nu\rho} = \hat \eta^{\mu}_{\phantom{\mu}\rho}
\eeq
In order to avoid potential ambiguities arising whenever Levi--Civita tensors $\varepsilon_{\mu\nu\rho}$ get contracted with $D$--dimensional objects\footnote{For a detailed discussion on this point in Chern--Simon--matter theories see for instance \cite{BGLP2,BGLP, GMPS}.}, we adopt the strategy to get rid of $\varepsilon$ tensors before promoting integrals to $D$ dimensions by using the following identity
\beq
\label{epsilon}
\varepsilon_{\lambda\mu\nu} \varepsilon_{\rho\sigma\tau} = \eta_{\l \rho} ( \eta_{\mu\s} \eta_{\nu\tau} - \eta_{\mu\tau} \eta_{\nu\s} ) 
- \eta_{\l \s} ( \eta_{\mu\rho} \eta_{\nu\tau} - \eta_{\mu\tau} \eta_{\nu\rho} ) + \eta_{\l \tau} ( \eta_{\mu\rho} \eta_{\nu\s} - \eta_{\mu\s} \eta_{\nu\rho} ) 
\eeq

When parametrizing the latitude by polar coordinates, the final contour integrals take the form of multiple integrations over $\epsilon$--dependent powers of trigonometric functions. 
We evaluate these integrals analytically by following the prescription of Refs. \cite{BGLP2, GMPS} to which we refer the reader for all the details. In the spirit of dimensional regularization we evaluate the integrals 
in regions of the $\epsilon$ parameter where they converge. We then rewrite the integrals as multiple series whose sum can be expressed in terms of hypergeometric functions. Finally, taking the $\epsilon \to 0$ limit requires performing a suitable analytic continuation of these functions close to the origin in the parameter space.  The results are expressed as an $\epsilon$--expansion up to finite terms.

We perform the calculation for $N,M$ finite (no large $N,M$ limit is taken). 

We stress that although the deformed Wilson loops depend in principle on two different parameters $\a$ and $\theta_0$, which have completely different origin, in the previous section they have been rephrased only in terms of the effective parameter $\nu$.  Therefore, we expect that also their quantum corrections will depend only on this combination.

\subsection{The one--loop result}

At one loop, contributions from purely bosonic diagrams are missing, since for a planar contour the graphs with the exchange of a gauge field are trivially zero. Therefore, we can immediately conclude that 
\beq
\langle W_B \rangle^{(1)} = 0 
\eeq
On the other hand, when we turn fermions on, a non--vanishing contribution arises from a fermion exchange diagram. It explicitly reads (in the following we always set $\ell=1$)
\begin{equation}
\langle W_F \rangle^{(1)} = -\frac{2\pi i{\cal R}}{k} \cos^2\theta_0 \, \frac{\Gamma\left( \frac32-\epsilon \right)}{2\pi^{\frac32-\epsilon}} \int_{\tau_1>\tau_2}\!\!\!\!\!\!\!\!
d\tau_1 d\tau_2
\left[ \frac{(\eta_1\gamma_\mu\bar\eta_2) x_{12}^\mu}{(x_{12}^2)^{3/2-\epsilon}}e^{-\frac{i\pi \nu}{2}}+
\frac{(\eta_2\gamma_\mu\bar\eta_1) x_{21}^\mu}{(x_{12}^2)^{3/2-\epsilon}}e^{\frac{i\pi \nu}{2}}\right]
\end{equation}
with the normalization factor ${\cal R}$ defined in \eqref{eq:T}.
Using the identities \eqref{Iden1} and \eqref{Iden2}  in appendix \ref{App:circle}, $\langle W_F \rangle^{(1)} $ can be rewritten as 
\begin{equation}\label{expression1234}
\langle W_F \rangle^{(1)} =
\frac{\pi i\, MN {\cal R}}{k} (2\cos{\theta_0})^{2\epsilon} \frac{\Gamma\left( \frac32-\epsilon \right)}{\pi^{\frac32-\epsilon}}
\, \left[\cos \left(\frac{\pi  \nu}{2}\right) (I_1 - I_2 \nu) - \sin \left(\frac{\pi  \nu}{2}\right) (I_3 + I_4 \nu)\right]
\end{equation}
where
\begin{subequations}
\bea
\displaystyle
I_1 = &\displaystyle\int_0^{2\pi}d\t_1 \int_0^{\t_2}d\t_2\, \cos \left(\frac{\t_{12}}{2}\right) \sin \left(\frac{\nu \t_{12}}{2}\right) \sin ^{2 \epsilon -2}\left(\frac{\t_{12}}{2}\right)
\\
\displaystyle I_2 =&\displaystyle\!\!\!\!\!\!\!\!\!\!\!\!\!\!\!\!\!\!\!\!\!\!\!\!\!\int_0^{2\pi}d\t_1 \int_0^{\t_2}d\t_2\, \cos \left( \frac{\nu\t_{12}}{2}\right) \sin ^{2 \epsilon -1}\left(\frac{\t_{12}}{2}\right)
\\
\displaystyle I_3 =&\displaystyle \int_0^{2\pi}d\t_1 \int_0^{\t_2}d\t_2\, \cos \left(\frac{\t_{12}}{2}\right) \cos \left( \frac{\nu\t_{12}}{2} \right) \sin ^{2 \epsilon -2}\left(\frac{\t_{12}}{2}\right)
\\
\displaystyle I_4 = &\!\!\!\!\!\!\!\!\!\!\!\!\!\!\!\!\!\!\!\!\!\!\!\!\!\!\!\!\!\!
\displaystyle\int_0^{2\pi}d\t_1 \int_0^{\t_2}d\t_2\, \sin \left(\frac{\nu \t_{12}}{2} \right) \sin ^{2 \epsilon -1}\left(\frac{\t_{12}}{2}\right)
\eea
\end{subequations}
The values of these integrals expanded around $\epsilon=0$ are given in appendix \ref{App:oneloop}. Short distance divergences appear in $I_1, I_2, I_3$ as simple poles in $\epsilon$, while $I_{4}$ happens to be finite. However it is easy to realize that in the combination (\ref{expression1234}) the divergent terms, as well as the special functions present in $I_{i}$,  completely  cancel. Inserting the explicit expression for ${\cal R}$, eq.  \eqref{eq:T}, the final result simply reads
\beq
\langle W_F \rangle^{(1)}  =  -\frac{MN}{k} \frac{2\pi l \nu}{(N+M) \tan{(\frac{\pi}{2}\nu)} + i(N-M)}  
\eeq
In the ABJM case ($M=N$) the result takes an even simpler form 
\begin{equation}
\langle W_F \rangle^{(1)} \Big|_{ABJM}  = -\frac{\pi\, N}{k} \, \nu \, \cot \left(\frac{\pi  \nu}{2}\right)
\end{equation}
As a check, we note that setting $\nu = 1$ ($\alpha = \tfrac{\pi}{4}, \th_0=0$) the result vanishes and we are back to the $1/2-$BPS case \cite{DT}.

A number of interesting observations are now in order. 
First of all, the generalized fermionic Wilson loop \eqref{eq:fermionic} is a new example in three dimensions where a non--trivial one--loop contribution arises, similarly to the case of the fermionic cusp discussed in \cite{Griguolo}. This contribution is generically complex and becomes real for $M=N$. The imaginary part, being proportional to $(N-M)$ is parity odd. 
 
Usually, in three dimensional Chern--Simons theories a non--vanishing, purely imaginary contribution at one--loop could signal the appearance of a non--trivial framing \cite{Witten, GMM}. However, in our case  we are not using a {\it contour splitting} regularization and the result should correspond to framing zero. Moreover, the contribution that we find is not purely imaginary. It is then interesting to understand which is its origin  in the present case. We will come back to this point in the last section where
we will argue that this factor may be interpreted  as the analogue of a non--integer framing.

\subsection{The two--loop result}\label{sec:two-loop}

We now discuss the two--loop corrections to the bosonic and fermionic Wilson loops.
The evaluation of two--loop diagrams, especially those involving fermionic contributions, turns out to be rather intricate. Therefore, 
we provide a detailed derivation of the bosonic diagrams only, deferring the discussion of fermionic graphs, including their regularization issues, to appendix \ref{app:fermionic}. 
 
\bigskip

\subsubsection*{Two loops: The bosonic diagrams}

\FIGURE{
 \includegraphics[width=0.21\textwidth]{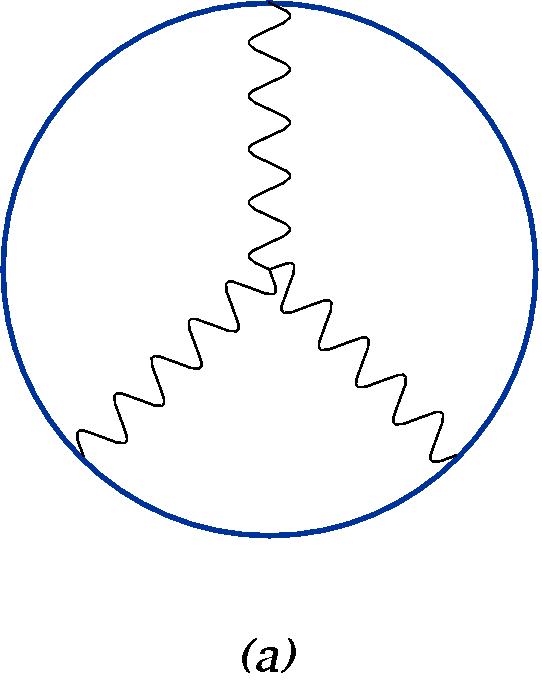}
    \caption{Pure Chern--Simons contribution.}
    \label{fig:vertexCS}
}

We first focus on diagrams emerging from the
 diagonal part of the superconnection
\eqref{supermatrix}, {\it i.e.} merely bosonic diagrams.
They contribute to the expectation value of both the bosonic \eqref{eq:bosonic} and the fermionic \eqref{eq:fermionic} Wilson loops, though with slightly different couplings ${\cal M}, \widehat{\cal M}$ to the scalar bilinears.

Their evaluation is straightforward and parallels the well--known computation of the $1/6-$BPS Wilson loop \cite{Drukker,Chen,Rey}. In dimensional regularization (then assuming framing zero), the only non--trivial contributions come from the three diagrams depicted in Figs. \ref{fig:vertexCS} and \ref{fig:bosonic}, as the rest of diagrams vanish due to symmetry arguments.

The only pure Chern--Simons contribution at two loops is associated to the vertex diagram of Figure \ref{fig:vertexCS} where the wavy lines correspond to the vector fields $A_\mu$ and $\hat{A}_\mu$. Focusing on the $A_\mu$--term, we have
\begin{align}
\label{int}
{\rm (a)} = -  \, \frac{N^3 e^{-i\frac{\pi}{2} \nu} {\cal R}}{k^2} \;
\frac{\G^3(\frac32 - \e)}{2\pi^{\frac{5}{2} - 3\e}}
\int d\tau_{1>2>3} \, \dot{x}_1^\s \, \dot{x}_2^\eta \, \dot{x}_3^\zeta \, \e^{\mu\nu\rho} \e_{\s\mu\xi} \e_{\eta\nu\tau} \e_{\zeta\rho \kappa}  
\,  I^{\xi \tau \kappa}  
\end{align}
where 
\begin{equation}
I^{\xi \tau \kappa} \equiv \int d^3x  \,  
\frac{  (x-x_1)^\xi (x-x_2)^\tau (x-x_3)^\kappa}{|x-x_1|^{3-2\e} |x-x_2|^{3-2\e} |x-x_3|^{3-2\e} } 
\end{equation}
This integral is well--known from Chern--Simons literature \cite{GMM} and, being finite, can be computed at $\e =0$, giving $ \frac{8}{3} \pi^3$. 
Hence, the contribution to the bosonic Wilson loop \eqref{eq:bosonic} from this diagrams reads
\begin{equation}
\label{eq:a1}
{\rm (a)}_B = - \frac{N^2-1}{k^2}  \, \frac{\pi^2}{6}
\end{equation}
By simply replacing $N$ with $M$ we obtain the contribution to the second  Wilson loop in \eqref{eq:bosonic}.

For the fermionic case we need to combine the two results in the supertrace of the superconnection multiplied by the matrix $\cal T$, according to \eqref{eq:fermionic}. We obtain
\beq
\label{eq:a2}
{\rm (a)}_F =   - \frac{(N^3-N) e^{-i\frac{\pi}{2} \nu}  - (M^3-M)e^{i\frac{\pi}{2} \nu} }{N  e^{-i\frac{\pi}{2} \nu} - M e^{i\frac{\pi}{2} \nu} }  \,  \frac{1}{k^2}  \, \frac{\pi^2}{6} 
\eeq
\FIGURE[l]{
 \includegraphics[width=0.40\textwidth, height=3.5cm]{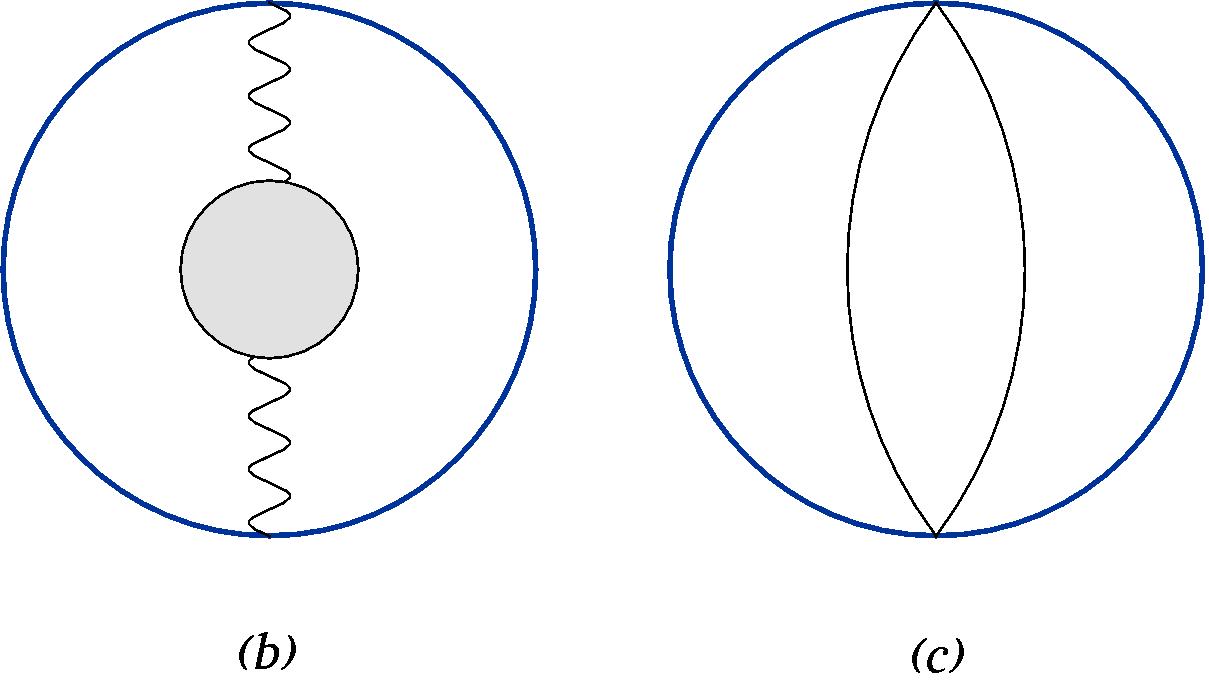}
    \caption{Combined bosonic contributions at two loops.}
    \label{fig:bosonic}
}
From the second order expansion of the exponentials of the two BPS Wilson loops  we obtain the diagrams in Fig. \ref{fig:bosonic}.
The former features the one--loop correction to the gauge propagator, which is non--vanishing thanks to matter fields running inside the loop, whereas the latter originates from the contractions of the scalar bilinears and is controlled by the matrix $\cal M$ ($\widehat{\cal M}$ in the bosonic case).
Such pieces can be conveniently combined into an effective contribution \cite{Drukker,Chen,Rey}.
For the upper-left part of the superconnection it reads
\begin{equation}
[{\rm (b) + (c)}] =     \,  \frac{MN^2 {\cal R}}{k^2} \, \frac{\G^2(\frac12-\e)}{\pi^{1 -2\e}} \; \int d\tau_{1>2} \, \frac{- \dot{x}_1 \cdot \dot{x}_2 + \frac{1}{4} \,  |\dot{x}_1| |\dot{x}_2| \, \Tr({\cal M}_1 {\cal M}_2)}{[(x_1-x_2)^2]^{1-2\e}}
\end{equation}
The same expression with $\Tr({\cal M}_1 {\cal M}_2)$ replaced by $\Tr(\widehat{\cal M}_1 \widehat{\cal M}_2)$ gives the contribution to the bosonic Wilson loop. However,
it is easy to realize that $\Tr({\cal M}_1{\cal M}_2) = \Tr(\widehat{\cal M}_1 \widehat{\cal M}_2)$, since the two matrices in \eqref{eq:matrixbosonic} and \eqref{eq:matrixfermionic} only differ by a sign in the last diagonal entry. Therefore, both contributions can be obtained by evaluating this integral with $\Tr({\cal M}_1 {\cal M}_2)$ explicitly given in eq.  \eqref{matrices}.

Setting ${\cal R} = 1/N$ the result gives the contribution to the bosonic $W_B$, whereas exchanging $N \leftrightarrow M$ we obtain the contribution to $\hat{W}_B$. For the fermionic one we have to combine the two results, take the supertrace with the matrix ${\cal T}$ and use the proper normalization \eqref{eq:T}. It turns out that, due to nice cancellations between ${\cal R}$ and factors in the numerator, the contributions for the bosonic and the fermionic Wilson loops are actually the same and read
\begin{equation}
\label{bcresult}
 [{\rm (b) + (c)}] =  \frac{NM}{k^2} \, \frac{\pi^2}{2} \, (1 + \nu^2)
\end{equation}

\bigskip

\subsubsection*{Two loops: The fermionic diagrams}

We now turn to the fermionic diagrams, which contribute exclusively to the fermionic latitude. They are depicted in Fig. \ref{fig:2loopdiagrams}. Details of the calculation are given in appendix D.
\FIGURE{
 \includegraphics[width=0.6\textwidth]{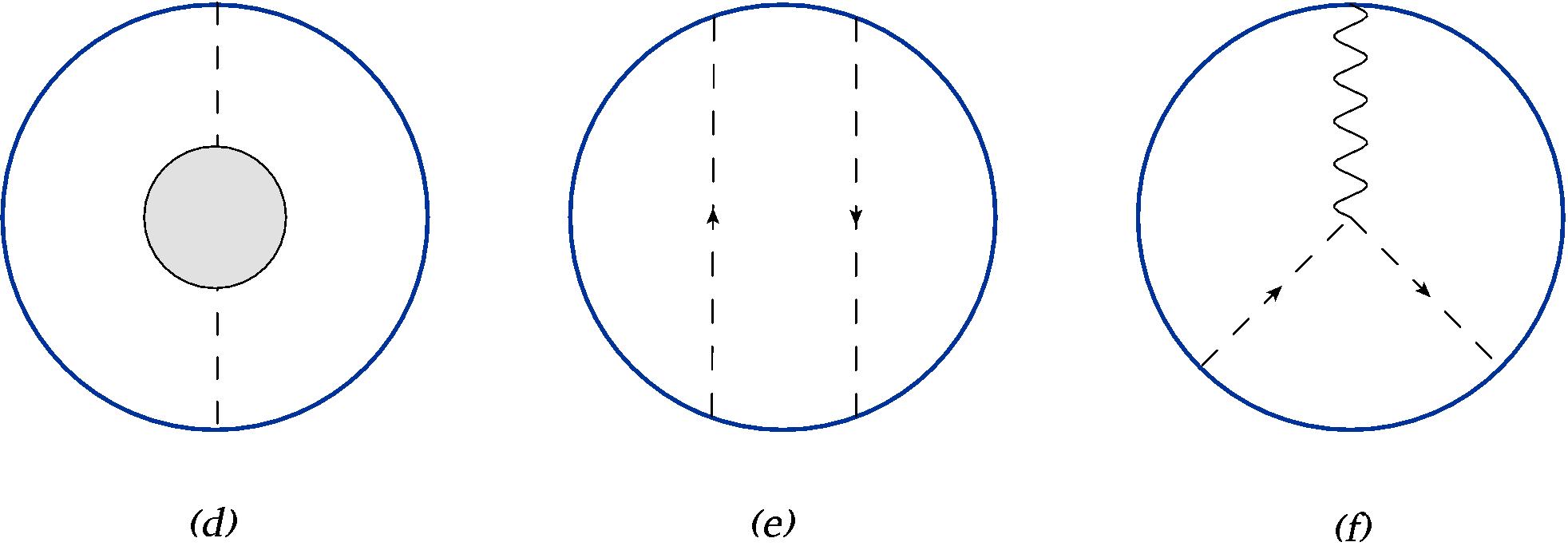}
    \caption{Fermionic contributions at two loops.}
    \label{fig:2loopdiagrams}
}

Diagram (d) emerges from the one--loop correction to the fermion propagator. While for the fermionic $1/2-$BPS Wilson loop such a contribution vanishes identically for symmetry reasons \cite{BGLP, BGLP3, GMPS}, its latitude deformation spoils those arguments and yields a finite result
\begin{equation}
\label{dresult}
{\rm (d)} =  {\cal R} \, (N-M) \, \frac{MN}{k^2} \, \pi \, \left( \nu - \frac{1}{\nu} \right) \sin{\frac{\pi \nu}{2}} 
\end{equation}

Diagram (e) accounts for the exchange of two fermions and corresponds to two possible contractions. Contrary to the $1/2-$BPS case it is divergent and henceforth contains a pole in the dimensional regularization parameter
\begin{align}
&{\rm (e)} = -{\cal R} MN \,(N + M)\, \left(\frac{2\pi}{k}\right)^2   
\frac{\Gamma^2(\tfrac{1}{2}-\epsilon)}{16 \pi^{3-2\epsilon}}   \, 
\frac{4i\pi\nu\cos(\tfrac{\pi\nu}{2})}{\epsilon} \,  (\cos \theta_0)^{4\e}
\nonumber\\
&+{\cal R}  \,  N M(N + M) \frac{i\pi\nu}{2 k^2} 
\left[ \pi(\nu-4)\sin(\tfrac{\pi\nu}{2})+8\cos(\tfrac{\pi\nu}{2})
\,  H_{\tfrac{\nu}{2} -\tfrac12} \right]
\non \\
&- {\cal R}  N M(M - N) \frac{\pi^2 \nu^2}{2k^2} \, \cos{\frac{\pi\nu}{2}}
\end{align}
Here $H_x$ stands for the harmonic numbers (see eq. \eqref{eq:harmonics}).

The last contribution from diagram (f) is the most involved. It requires special care in dealing with the antisymmetric Levi--Civita tensors in the presence of dimensional regularization, as outlined at the beginning of this section. Moreover it requires solving a space-time integral in dimensional regularization. This is a hard task that can be accomplished via a delicate subtraction of the divergence, according to the prescription outlined in \cite{GMPS}. The integrals on the loop parameters are themselves rather complicated and are solved separately in appendix \ref{app:fermionic}. The final result for this diagram reads
\begin{align}
\label{firstcontracted2}
{\rm (f)}&= {\cal R}\, N M(M + N)  \left(\frac{2\pi}{k}\right)^2\, \frac{\Gamma^2( \tfrac12-\epsilon)}{16 \pi^{3-2\epsilon}}  \, \frac{4 i \pi \nu \cos \left(\frac{\pi  \nu }{2}\right)}{\e} \, (\cos{\th})^{4\e} 
\non \\
&~ + {\cal R} \, MN (M + N) \, \frac{i\pi\nu}{2k^2} \left[  4\pi \sin \left(\frac{\pi  \nu }{2}\right) - 8\cos \left(\frac{\pi  \nu }{2}\right) \,  H_{\tfrac{\nu}{2} -\tfrac12} \right]
\nonumber \\
&~ - {\cal R} \,  (N -  M)  \frac{MN}{k^2} \, \pi \left( \nu - \frac{1}{\nu} \right) \sin \left(\frac{\pi  \nu }{2}\right)
\end{align}

\bigskip
\subsubsection*{Result for the fermionic Wilson loop}

Combining results \eqref{eq:a2} and \eqref{bcresult} with the contributions from the fermionic diagrams we obtain the complete two--loop expectation value for the deformed 
fermionic Wilson loop.

Short distance divergences appearing  in diagrams (e) and (f)  cancel each other, so leading to a finite result. This is  consistent with the fact that our Wilson loop partially preserves supersymmetry. Furthermore, the contribution from diagram (d) is completely cancelled by an opposite term in \eqref{firstcontracted2} from diagram (f). Therefore, the complete fermionic sector collapses to the following simple form
\begin{equation}
\label{eq:resultdef}
(d) + (e)+(f) 
 = \frac{\pi ^2 \nu ^2 M N}{2 k^2} \, \frac{N e^{i\frac{\pi  \nu }{2}} - M e^{-i\frac{\pi  \nu }{2}}}{N e^{-i\frac{\pi  \nu }{2}} - M e^{i\frac{\pi  \nu }{2}}}
\end{equation}
Adding the contributions (\ref{eq:a2}, \ref{bcresult}) and taking into account the one--loop result, we finally have
\begin{framed}
\begin{align}
\label{eq:Wffinal}
& \langle W_F(\nu) \rangle
 =  1 + \frac{2\pi i\, M N  {\cal R}}{k} \, \nu \, \cos \frac{\pi  \nu}{2} 
\non \\
& \qquad \qquad \quad -  \frac{\pi^2 {\cal R}}{6k^2} \, \left[ (N-M)(N^2+M^2 +2(\nu^2-1)MN-1) \cos{\frac{\pi \nu}{2}} 
\right.\non \\&~ \left. \qquad \qquad \quad \qquad 
- i (N+M) (N^2+M^2 -4MN-1)  \sin{\frac{\pi \nu}{2}} \right] + {\cal O}\left(k^{-3}\right)
\end{align} 
\end{framed}
\bigskip
In the ABJM case it reduces to 
\beq
\label{WLABJM}
\langle W_F(\nu) \rangle_{ABJM}  
= 1 - \frac{N}{k} \, \pi  \, \nu \, {\rm ctg}{\frac{\pi \nu}{2}} +\frac{N^2}{k^2} \frac{\pi^2}{6} \left(2+\frac{1}{N^2}\right) 
 + {\cal O}\left(k^{-3}\right)
\eeq
Quite remarkably, the $\nu$--dependence survives only in the one--loop contribution. This has nice implications for the connection of our results with the Bremsstrahlung function, as we are going to discuss in the next section. 
 
Interesting limits of the result \eqref{eq:Wffinal} are $\nu \to 1,0$. For $\nu=1$ it reduces to the two--loop expression for the $1/2-$BPS Wilson loop on the maximal circle \cite{Drukker}.
Instead for $\nu \to 0$ the deformed fermionic Wilson loop reduces to a Zarembo-like operator \cite{zarembo}, belonging to a family of operators preserving Poincar\'e supercharges only \cite{Cardinali}.  We obtain 
\beq
\label{eq:zarembo}
\lim_{\nu \to 0}  \langle W_F (\nu)\rangle 
= 1 - \frac{\pi^2}{6k^2} \, \left[ (N-M)^2 -1 \right]  + {\cal O}\left(k^{-3}\right)
\eeq  
which coincides with the two--loop result for a bosonic Wilson loop in pure Chern--Simons theory with gauge group $U(|N-M|)$ \cite{Witten}.

\subsubsection*{Result for the bosonic Wilson loop}

Combining the results \eqref{eq:a1} and \eqref{bcresult} from the bosonic diagrams we obtain the two--loop expectation value of Wilson loops \eqref{eq:bosonic}
\begin{framed}
\begin{align}
\label{bosonicW}
\langle W_B (\nu)\rangle
&= 1 + \frac{\pi^2}{k^2} \, \left[ \frac12 (1 + \nu^2) MN - \frac{N^2-1}{6} \right] + {\cal O}\left(k^{-3}\right) 
\non \\
\langle \hat{W}_B (\nu)\rangle
&= 1 + \frac{\pi^2}{k^2} \, \left[ \frac12 (1 + \nu^2) MN - \frac{M^2-1}{6} \right] + {\cal O}\left(k^{-3}\right)
\end{align}
\end{framed}
They provide the two--loop result for a latitude, bosonic $1/12-$BPS Wilson loop and for the ABJM theory ($M=N$) they coincide. As expected, for $\nu =1$ the results \eqref{bosonicW} reproduce the corresponding expressions for the $1/6-$BPS Wilson loops on the maximal circle with ${\cal M} = {\rm diag} (-1,1,-1,1)$ \cite{Drukker, Rey, BGLP3}.

\bigskip

\section{Discussion}
\label{sec:discussion}
 
The explicit evaluation we have performed in the previous section provides important information about the general structure of the full quantum result. Moreover, in the ABJM case it suggests some interesting relations with the so-called Bremsstrahlung function. Here we elaborate on these aspects. 

\subsection{Non--integer framing}

A first crucial point that we are going to discuss is how the equivalence between the fermionic Wilson loops (\ref{supermatrix}, \ref{eq:matrixfermionic}) and the bosonic loops (\ref{QL3}, \ref{eq:matrixbosonic}), as expressed by the cohomological relation (\ref{QL10}) gets implemented at quantum level. 

In the undeformed circular case ($\nu=1$) it has been observed \cite{DT,Drukker:2010nc} that the quantum realization of the cohomological equivalence involves a particular choice of framing in the actual computation. More precisely, the expected relation
\beq
\label{QL4}
\langle W_F(1)\rangle_1 = \frac{N\, \langle W_B(1)\rangle_1 + M\, \langle\hat W_B(1)\rangle_1}{N +M}
\eeq
is obtained only when the vacuum expectation value is computed at framing one\footnote{This is the meaning of the subscript 1.}. The results obtained by means of localization techniques, $i.\,e.$ through averages in the relevant matrix models \cite{Kapustin}, display this feature clearly. On the other hand, conventional perturbation theory where diagrams are evaluated in DRED regularization leads to results at framing zero. The appropriate relation is then modified as
\beq
\label{QL5}
\langle W_F(1)\rangle_0 = e^{-\frac{i\pi(N-M)}{k}} \;  \frac{N\, e^{\frac{i\pi  N}{k}}\,\langle W_B(1)\rangle_0 + M\, e^{-\frac{i\pi  M}{k}}\, \langle\hat W_B(1)\rangle_0}{N +M}
\eeq
This expression has been thoroughly checked at two--loop level in \cite{BGLP, BGLP3, GMPS}. 

In the present case we expect an analogous situation, with some phase factors correcting (\ref{QL3}) when the fermionic and the bosonic Wilson loops are computed in perturbation theory. From the direct inspection of (\ref{eq:Wffinal}) and (\ref{bosonicW}) we obtain (up to second order in $1/k$)
\beq
\label{QL6}
\langle W_F(\nu)\rangle_0={\cal R}\, e^{-\frac{i\pi \nu (N-M)}{k}} \left[N\, e^{-\frac{i\pi \nu}{2}}\,e^{\frac{i\pi \nu N}{k}}\, \langle W_B(\nu)\rangle_0 -M\, e^{\frac{i\pi \nu} {2}}\,e^{-\frac{i\pi\nu  M}{k}}\, \langle\hat W_B(\nu)\rangle_0\right]
\eeq

On the other hand any regularization exactly compatible with the cohomological equivalence between fermionic and bosonic Wilson loops should respect (\ref{QL3},\ref{QL10}). Inspired by our two--loop computation and the analogy with the undeformed circular case, we are led to conjecture that the correct relation between the perturbative (``zero-framing'') result and a putative ``framed" computation, consistent with the ${\cal Q}$-exactness of the fermionic couplings, should be obtained by defining ``framing $\nu$'' quantities
\begin{align}
\label{bosonicWfram}
& \langle W_B (\nu)\rangle_\nu \equiv   e^{\frac{i\pi\nu  N}{k}}\,\langle\hat W_B(\nu)\rangle_0  \quad , \quad 
\langle \hat{W}_B (\nu)\rangle_\nu  \equiv e^{-\frac{i\pi\nu  M}{k}}\,\langle\hat W_B(\nu)\rangle_0
\non \\
& \langle W_F(\nu)\rangle_\nu \equiv e^{\frac{i\pi\nu  (N-M)}{k}}\,\langle W_F(\nu)\rangle_0
\end{align}
that reproduce
\begin{framed}
\beq
\label{equivalence}
\langle W_F(\nu)\rangle_\nu={\cal R}\, \left[N\, e^{-\frac{i\pi \nu}{2}}\, \langle W_B(\nu)\rangle_\nu -M\, e^{\frac{i\pi \nu} {2}}\, \langle\hat W_B(\nu)\rangle_\nu\right]
\eeq
\end{framed}
Strictly speaking, framing is a topological property and it should be parametrized by integer numbers \cite{Witten, GMM}, so our proposal may sound somehow paradoxical. However, we recall that we are not in a topological theory and a framing procedure could produce an explicit dependence of the phase on the length of the contour and the matter couplings. In fact, cases in which regularization techniques lead to similar dependences in Wilson loop computations are already present in the literature, even for pure Chern--Simons theories. For instance, in \cite{Hama:2011ea} it has been shown that $1/2-$BPS loops on a squashed $S^3$ can be evaluated by a straightforward application of supersymmetric localization in pure Chern--Simons theory. In particular one can define two different unknot $1/2-$BPS operators and the computation gives the expected (topological invariant) result, up to an overall phase 
\beq
\exp{\frac{i\pi b  N}{k}}
\eeq
 where $b$ is the squashing parameter\footnote{Actually this is the result for one class of $1/2-$BPS loops. In the other case the framing phase is obtained by sending $b \to  b^{-1}$.}. 

\noindent
Another surprising feature of our results concerns the limit $\nu\to 0$. This corresponds to a Zarembo-like circle (for $\a=0$) or, equivalently, to a vanishing latitude shrinking on the north pole (for $\th_0=\pi/2$). In both cases we would have expected a decoupling of the matter contributions and the recovery of the pure Chern--Simons vacuum expectation value. Instead, we observe that in this limit a residual presence of the matter loops changes the topological result, which seems to reduce to a Chern--Simons average in $U(|N-M|)$ theory, at least up to two loops (see eq \eqref{eq:zarembo}). 
This pattern is not present in ${\cal N}=4$ SYM where in both limits a trivial observable is recovered. At the moment we do not have a general explanation of the appearance of this effect in three dimensions, neither we can assure that it will persist at higher loops.

\subsection{ABJM Bremsstrahlung function from the deformed circle}\label{sec:brem}

We now discuss the implications of our results on the study of the Bremsstrahlung function $B(\lambda,N)$ in ABJM theory. From the physical point of view this quantity determines the energy emitted by a moving quark in the small velocity limit

\beq
\Delta E=2 \pi B\int dt\, (\dot v)^2
\eeq
In any conformal field theory it can be conveniently computed by means of a well--known observable that plays an ubiquitous role in quantum gauge theories, the so-called cusp anomalous dimension $\G_{cusp}(\varphi)$ \cite{Polyakov:1980ca}. In fact this quantity, which governs the singular behaviour of a Wilson operator close to a $\varphi$--cusp 
\beq
\langle W \rangle \simeq e^{-\G_{cusp}(\varphi)\, \log{(\L/\e)}}
\eeq
exhibits an interesting relation with $B$ in its small angle expansion \cite{Correa:2012at}
\beq
\G_{cusp}(\varphi)\simeq -B(\lambda,N)\, \varphi^2 \quad , \quad \varphi \ll 1
\eeq
Therefore, there is a strict relationship between Wilson loops, cusp anomalous dimensions and Bremsstrahlung function and the actual evaluation of one of them can in principle provide information on the other two quantities. 
 
In ${\cal N}=4$ SYM it was shown that $B(\lambda,N)$, although being not a BPS quantity, can be computed exactly \cite{Correa:2012at} at all values of the coupling and for all $N$, by using an approach based on supersymmetric localization. The result has been checked at weak \cite{Correa:2012nk, Drukker:2011za} and strong coupling \cite{Drukker:2011za,Forini:2010ek}. 
Remarkably, it can be also obtained solving in a suitable limit a TBA system of integral equations \cite{Correa:2012hh,Drukker:2012de} that extend  the original bulk system introduced in \cite{Bombardelli:2009ns,Gromov:2009bc,Arutyunov:2009ur} to a case with boundary. In this way results obtained using integrability are connected with results obtained using localization.

One of the main ingredients in the exact computation of $B(\lambda,N$) is the explicit non--perturbative expression for $1/8-$BPS Wilson loops on the $S^2$ sphere (a remarkable subclass of DGRT loops \cite{Drukker:2007dw, Drukker:2007qr, Drukker:2007yx}, that are computed by two--dimensional Yang--Mills theory in the zero--instanton sector \cite{Drukker:2007yx}). In particular, it takes advantage of the simple formula for a latitude loop $\langle W(\theta_0)\rangle$ in terms of Laguerre polynomials \cite{Drukker:2000rr,Drukker:2006ga} to compute the exact Bremsstrahlung function directly as \cite{Correa:2012at}
\beq\label{Bmalda}
B(\lambda,N) = -\frac{1}{4\pi^2}\, \frac{1}{\langle W(0)\rangle}\left.\partial^2_{\theta_0}\, \langle W(\theta_0)\rangle\right|_{\theta_0=0}
\eeq
where $\th_0$ is the internal latitude angle. As discussed in \cite{Drukker:2007dw, Drukker:2007qr, Drukker:2007yx}, the internal latitude angle $\theta_0$ can be interpreted as induced by a geometrical latitude loop angle. We remark that the derivation of this equation heavily relies on the BPS condition for the generalized cusp $\G_{cusp}(\varphi = \th) = 0$, which implies that for small angles  $\G_{cusp}(\varphi, \th)\simeq B(\lambda,N)(\th^2 - \varphi^2)$ \cite{Correa:2012at}.

For three dimensional superconformal theories, an analogous expression for $B(\lambda,N)$ obtained from first principles is still missing, although great progress has been recently done in \cite{Lewkowycz:2013laa} where a proposal for computing the Bremsstrahlung function from Wilson loops on a squashed sphere has been given. Here we further elaborate on this problem by exploiting the results we have obtained for fermionic and bosonic latitude operators. 

First of all, in ABJM theory an explicit result for the cusp anomalous dimension for fermionic Wilson operators is available. Precisely, the relevant calculation up to two loops has been performed in \cite{Griguolo} and in the planar limit it reads
\beq
\label{cuspGS}
\Gamma_{1/2}[\varphi,\th]=-\l \, \left[\frac{\cos\th/2}{\cos\varphi/2}-1  \right]- \l^2 \, \left[\frac{\cos\th/2}{\cos\varphi/2}-1  \right]\log^2 \left(\cos\varphi/2\right)
\eeq 
This result depends on the geometric angle $\varphi$ and an internal angle $\th$ measuring the relative $R$-symmetry orientation on the two halves of the cusp\footnote{ The scalar and fermionic couplings on the two halves are chosen so that the cusp is locally $1/2-$BPS.}. When $ \varphi= \pm \th$ the configuration is supersymmetric and therefore the cusp anomalous dimension vanishes. Like for ${\cal N}=4$ SYM case, the behaviour of the generalized cusp for small angles is then $\G_{1/2}(\varphi, \th)\simeq B(\lambda)(\th^2 - \varphi^2)$. 
Therefore, we can extract the Bremsstrahlung function (at zero cusp angle) from  
\beq
B(\lambda)=- \frac{1}{2}\, \partial^2_\varphi\, \Gamma[\varphi,0]\,\Big|_{\varphi=0}
\eeq
which applied to \eqref{cuspGS} provides the two--loop result
\beq\label{B1/2}
B_{1/2}(\lambda)=\frac{\lambda}{8} + {\cal O}\left(\lambda^3\right)
\eeq  
We observe that, quite surprisingly, we do not have a contribution proportional to $\lambda^2$, in spite of the non--trivial form of $\G_{1/2}$. 

We can now check whether applying a prescription similar to \eqref{Bmalda} to our fermionic latitude we reproduce expression \eqref{B1/2}. 

Our result \eqref{WLABJM} involves a geometric latitude angle $\th_0$ on $S^2$ and an internal angle $\a$ in the space of the $SU(4)$ couplings, which however combine in a single parameter $\nu= \sin{2\a} \cos{\th_0}$.  Trading derivatives respect to the $\theta_0$ angle for derivatives with respect to $\nu$ ($\nu=1$ for $\th_0=0$ and $\a = \pi/4$)
\beq
\frac{\partial^2}{\partial{\theta_0^2}}\, \Big|_{\theta_0=0, \a = \pi/4 }=-\frac{\partial}{\partial\nu}\, \Big|_{\nu=1} 
\eeq
we find that the prescription 
\begin{framed}
\beq
\label{conjecture}
B_{1/2} (\lambda) = \frac{1}{4\pi^2}\, \partial_\nu\,\log\, \langle W_F(\nu)\rangle_0 ~ \Big|_{\nu=1}
\eeq
\end{framed}
\noindent
applied to our two--loop result \eqref{WLABJM} reproduces exactly eq. \eqref{B1/2}. In particular, it is quite remarkable that for the ABJM case the perturbative result for the latitude Wilson loop looses the $\nu$ dependence at two loops, so leading to perfect consistency with the absence of a $\lambda^2$ term in $B_{1/2}(\l)$ as obtained from the generalized cusp.   

Therefore, in analogy with ${\cal N}=4$ SYM and supported by our two--loop explicit check, for the ABJM theory we propose eq. \eqref{conjecture} as the prescription for computing the exact Bremsstrahlung function from the fermionic Wilson loop.  

It is interesting to rewrite the above equation using the relation \eqref{equivalence} between fermionic and bosonic Wilson loops. It is not difficult to obtain
\bea
\label{Bfin}
B_{1/2}(\lambda)&=& \frac{1}{4\pi^2}\left[\left.\partial_\nu\, \log\left(\langle W_B(\nu)\rangle_\nu+\langle \hat W_B(\nu)\rangle_\nu\right)\right|_{\nu=1}-\frac{i\pi}{2}\, \frac{\langle W_B(1)\rangle_1-\langle \hat W_B(1)\rangle_1}{\langle W_B(1)\rangle_1+\langle \hat W_B(1)\rangle_1}\right]
\non \\
&=& \frac{1}{4\pi^2}\left[\left.\partial_\nu\,\log\left(\langle W_B(\nu)\rangle_\nu+\langle \hat W_B(\nu)\rangle_\nu\right)\right|_{\nu=1}+\frac{\pi}{2} \, {\rm tg}\Phi_B\right]
\eea
where in the second line we have used the relation $\langle \hat W_B(\nu)\rangle_1=\langle  W_B(\nu)\rangle_1^*$ to express the result in terms of the phase $\Phi_B$ of the $1/6-$BPS bosonic loop on the maximal circle. In particular, our proposal always leads to a manifestly real Bremsstrahlung function (in contrast with the proposal of \cite{Marmiroli}).  

Since $\langle \hat W_B\rangle_1(\l) =\langle  W_B\rangle_1(-\l)$, the first term in \eqref{Bfin}, when expanded, leads to even powers of $\l$, whereas the second term encodes all the odd powers. In particular, at the order we are working the result \eqref{B1/2} originates entirely from $\Phi_B$.
 
More generally, since the second term can be easily evaluated both at weak and strong coupling by means of the results already available for the undeformed bosonic Wilson loop \cite{Kapustin, Klemm:2012ii}, we can make a prediction for the three loop contribution to $B_{1/2}(\lambda)$, that could be tested by an explicit computation of the cusp anomalous dimension in ABJM theory at that order. Taking into account the series expansion for the $1/6-$BPS Wilson loop given in \cite{Kapustin,Drukker:2010nc} we obtain
\beq
\label{prediction}
B_{1/2}(\lambda)=\frac{\lambda}{8}-\frac{\pi^2}{48}\lambda^3 + {\cal O}\left(\lambda^5\right)
\eeq

For the full understanding of (\ref{Bfin}) it would be necessary to know the exact expression for $\langle W_B(\nu)\rangle_\nu$ for generic values of $\nu$, or better its derivative with respect to $\nu$ evaluated at $\nu=1$. Unfortunately, this is still missing in the literature. 

A similar problem has been discussed very recently in \cite{Lewkowycz:2013laa} for the effective calculation of the Bremsstrahlung function associated to the 1/6 cusp. There, the derivative of the bosonic Wilson loop on a squashed sphere $S^3_b$, with respect to the squashing parameter $b$ has been shown to be relevant for the explicit computation, when evaluated at $b=1$. The authors argued that the same result can be obtained by considering a $n$-winding Wilson loop on the undeformed sphere and performing the derivative with respect to $n$, at $n=1$. The justification of this correspondence relies on the Renyi entropy as obtained from a matrix model on a $n$--branched sphere \cite{Nishioka:2013haa}, the explicit knowledge of the $b$ dependence in the complicated matrix-model encoding the Wilson loop average \cite{Hama:2011ea} and the equivalence between the matrix model on the squashed and the branched spheres under the identification $b=\sqrt{n}$. 

In our case it is tempting to propose a similar recipe, in spite of the fact that we do not have a solid argument to justify it (we do not have an expression for the matrix-model, if any, nor deep information about the behaviour of $\langle W_B(\nu)\rangle_\nu$  near $\nu=1$). Nevertheless, we assume that  
\beq\label{crazy}
\left. \partial_\nu\, \log\left(\langle W_B(\nu)\rangle_\nu+\langle \hat W_B(\nu)\rangle_\nu\right)\right|_{\nu=1}=\left. \partial_n\, \log\left(\langle W_n^{1/6}\rangle + \langle \hat W_n^{1/6}\rangle\right) \, \frac{\partial n(\nu)}{\partial \nu}\right|_{\nu=1}
\eeq
with $\langle W_n^{1/6}\rangle$ being the multiply wound Wilson loop on the great circle, $n=n(\nu)$ and $n(1)=1$. The expectation value $\langle W_n^{1/6}\rangle$ is known exactly from localization and comes with a crucial framing phase factor which depends on $n$. We can use this result to test the output of \eqref{crazy} against the existing data, both at weak and strong coupling.

First of all,  at weak coupling we should recover the two--loop result \eqref{B1/2} when we apply the prescription (\ref{crazy}) to the closed formula of \cite{Klemm:2012ii} for the eigenvalue density $\rho(\mu)$ in the 1/6 case. The $n$--winding loop is computed as the matrix-model average
\beq
\langle W_n^{1/6} \rangle = \int_{{\cal C}}\,d\mu\, \rho(\mu)\,\exp(n\mu)
\eeq
and the result at third order in the $\lambda$ expansion reads
\bea
\langle W_n^{1/6} \rangle &=& 1+i\pi n^2\lambda+\frac{\pi^2}{3}\left(2n^2-n^4\right)\lambda^2-i \frac{\pi^3}{18} \left(4n^2 - 8n^4 + n^6\right) \l^3 + {\cal O}\left(\lambda^4\right)\nonumber\\
\langle \hat W_n^{1/6} \rangle &=& \langle W_n^{1/6} \rangle^*
\eea
We find that at this order
\beq
\label{null}
\left.
\partial_n\, \log\left(\langle W_n^{1/6}\rangle+\langle \hat W_n^{1/6}\rangle \right)\right|_{n=1}=0
 \eeq
and no choice of $n(\nu)$ is needed in order to implement \eqref{crazy}. Rather, the perturbative prediction \eqref{prediction} comes entirely from the second term in \eqref{Bfin}.

More generally, using the weak coupling expansion of $\langle W_n^{1/6} \rangle$ given in appendix \ref{sec:expansions} it is easy to realize that this pattern persists at higher orders. This seems to suggest that at weak coupling $B_{1/2}(\l)$ should be described by an odd function of $\l$, as already observed in \cite{Lewkowycz:2013laa}. 

Another crucial test of our proposal \eqref{crazy} comes from comparing our formula with the strong coupling calculation of the cusp anomalous dimension in ABJM theory at small cusp angle \cite{Forini:2012bb}.  The exact expression found in \cite{Klemm:2012ii}, once expanded at strong coupling gives
\bea\label{KM}
\langle W_n^{1/6} \rangle &=& {\rm e}^{\frac{i\pi n}{2}}\left[\frac{\sqrt{2\lambda}}{4\pi n}-\left (\frac{H_n}{4\pi^2 n}+\frac{i}{8\pi n}+\frac{1}{96}\right) + \left( \frac{i}{192} + \frac{\pi n}{4608} + \frac{H_{n-1}}{96\pi} \right) \frac{1}{\sqrt{2\l}} \right.
\non \\
&~&\left . ~~~- \left( \frac{i\pi n}{18432} + \frac{\pi^2 n^2}{663552} + \frac{n H_{n-1}}{9216} \right) \frac{1}{\l} + {\cal O}\left(\lambda^{-\frac32}\right)\right]
{e}^{\pi n \sqrt{2\lambda}}
\eea
Applying the prescription (\ref{crazy}) to this expression we find that the result \eqref{null} is true also at strong coupling. Therefore, we do not need to guess the explicit function $n(\nu)$ and the non--trivial contribution to the Bremsstrahlung function comes once again only from the phase term in eq. \eqref{Bfin}. Explicitly we find 
\beq
\label{strong}
B_{1/2}(\lambda)=\frac{\sqrt{2\lambda}}{4\pi}-\frac{1}{4\pi^2} - \frac{1}{96\pi} \frac{1}{\sqrt{2\l}} + {\cal O} \left( \l^{-1} \right)
\eeq
Comparing this expression with the Bremsstrahlung function obtained from the explicit string computation performed in \cite{Forini:2012bb}, we find perfect agreement at leading order\footnote{We thank Valentina Forini for pointing out a factor $\tfrac12$ missing in the leading term of their result in \cite{Forini:2012bb}.} while a mismatch appears in the subleading term. The first contribution in (\ref{strong}) should correspond to the action of the classical string worldsheet, while the second order is obtained by evaluating the quantum fluctuations around the classical solution in the relevant sigma-model. This is a very complicated calculation that was indeed attempted in \cite{Forini:2012bb}, considering both a geometric (cusp) angle and an internal (R-symmetry) angle. As remarked by the authors, the final answer does not appear to respect the BPS condition, a fact that casts some doubts on the correctness of the relevant coefficient in our comparison.

In ABJM theory integrability is a powerful tool to get exact results \cite{oai:arXiv.org:0806.3951}-\cite{oai:arXiv.org:0807.0777}, very much as in ${\cal N}=4$ SYM. However, in the three dimensional case there is still one player missing in the integrability game: the infamous function $h(\lambda)$, mastering the dispersion relation of a single magnon moving on the spin chain \cite{oai:arXiv.org:0806.3391,oai:arXiv.org:0806.4589,oai:arXiv.org:0806.4959}. Weak \cite{oai:arXiv.org:0908.2463,Leoni:2010tb} and strong \cite{oai:arXiv.org:0809.4038} coupling expressions have been obtained at leading and subleading orders, but no systematic method for computing $h(\lambda)$ exists yet. On the other hand, it should be possible to find a three dimensional analogue of the set of TBA integral equations, used in \cite{Correa:2012hh,Drukker:2012de}, and apply it to the actual computation of $B(\lambda)$, as done in \cite{Gromov:2012eu,Gromov:2013qga}.  Remarkably, a three dimensional set of TBA equations describing the bulk system has been discovered and studied in \cite{Bombardelli:2009xz,Gromov:2009at,Cavaglia':2013hva}. Having in this calculation $h(\lambda)$ as input, a direct comparison with our proposal for $B(\l)$ would provide, in principle, an all--order definition for $h$.

\vskip 15pt

We close this section with few remarks about the case of bosonic latitude Wilson loops. 
In principle, we can use our result \eqref{bosonicW} for bosonic latitude loops to test recent proposals appeared in the literature \cite{Lewkowycz:2013laa} for the Bremsstrahlung function related to $1/6-$BPS cusps\footnote{By 
$1/6-$BPS cusp we mean that the lines forming the cusp are locally $1/6-$BPS.}. For a purely bosonic cusp, the direct computation performed in \cite{Griguolo} gives
\beq\label{cuspb}
\Gamma_{1/6}[\varphi,\th] \simeq \l^2 \left[\cos\varphi-\cos^2\frac{\th}{2}\right] \simeq -\frac{\lambda^2}{2}\left(\varphi^2-\frac{\th^2}{2}\right)  \  \  \ {\rm for} \  \  \varphi, \th \ll 1
\eeq
We remark that this cusp is not BPS at $\varphi=\pm \th$, except that in the particular case $\varphi=\th=0$. 

Taking the second derivative with respect to $\varphi$, or simply looking at  the coefficient of $\varphi^2$, we get
\beq
B_{1/6}(\lambda)=\frac{\lambda^2}{2} + {\cal O}\left(\lambda^3\right)
\eeq
that at this order coincides with the results obtained in \cite{Lewkowycz:2013laa}. 

We can study whether this expression can be extracted directly from our explicit result for the two--loop bosonic latitude, eq. \eqref{bosonicW}, by applying a prescription similar to \eqref{conjecture}. Formally writing  
\beq
B_{1/6}(\lambda)= \frac{1}{4\pi^2}\, \partial_\nu\,\log\, \langle W_B(\nu)\rangle_0\, \Big|_{\nu=1}
\eeq
we obtain  
\beq
B_{1/6}(\lambda)= \frac{\lambda^2}{4} + {\cal O}\left(\lambda^3\right)
\eeq
Apparently in this case we have a mismatch of a factor 1/2. However, we recall that the derivation of the function $B$ from the latitude Wilson loop given in \cite{Correa:2012at} assumes the relevant cusp to be BPS at $\varphi=\th$, allowing to identify the Bremsstrahlung term with the coefficient of $\th^2$, in the limit $\varphi, \th \to 0$. This is not what happens in the present case, rather the coefficients of $\varphi^2$ and $\th^2$ in \eqref{cuspb} differ by a factor $1/2$. This explains the apparent mismatch that we observe.

\section*{Acknowledgements}

We thank Jeremias Aguilera Damia, Diego Correa, Davide Fioravanti, Valentina Forini, Gaston Giribet, Aitor Lewkowycz, Juan Maldacena, Daniele Marmiroli, Gabriele Martelloni and Guillermo Silva for very useful discussions. 
The work of MB has been supported by the Volkswagen-Foundation.
The work of ML has been supported by the research project CONICET PIP0396.
This work has been supported in part by INFN and COST Action MP1210 "The String Theory Universe".

\vfill
\newpage

\appendix

\section{Conventions and Feynman rules}\label{sec:conventions}

Along the paper we have strictly stuck to conventions of Ref. \cite{BGLP3}. In order to facilitate the reading we give a brief summary of the main ones.

We work in euclidean three--dimensional space with coordinates $x^\mu = (x^0, x^1, x^2)$. We choose a set of gamma matrices satisfying Clifford algebra $\{ \g^\mu , \g^\nu \} = 2 \d^{\mu\nu} \mathbb{I}$ as
\beq
(\g^\mu)_\a^{\; \, \b} = \{ -\s^3, \s^1, \s^2 \}
\eeq
Useful identities are
\bea 
&&  \g^\mu \g^\nu = \d^{\mu \nu} \mathbb{I} - i \varepsilon^{\mu\nu\rho} \g^\rho
\non \\
&& \g^\mu \g^\nu \g^\rho = \d^{\mu\nu} \g^\rho - \d^{\mu\rho} \g^\nu+  \d^{\nu\rho} \g^\mu  - i \varepsilon^{\mu\nu\rho} \mathbb{I}
\non \\
&& 
\g^\mu \g^\nu \g^\rho \g^\s -  \g^\s \g^\rho \g^\nu \g^\mu = -2i \left( \d^{\mu\nu} \varepsilon^{\rho\s \eta}  + \d^{\rho \s}  \varepsilon^{\mu\nu\eta} + \d^{\nu\eta} \varepsilon^{\rho \mu \s} +
\d^{\mu\eta} \varepsilon^{\nu\rho\s}  \right) \g^\eta
 \\
&& 
\non \\
&&
\Tr (\g^\mu \g^\nu) = 2 \d^{\mu\nu}
\non \\
&&
\Tr (\g^\mu \g^\nu \g^\rho) = -2i \varepsilon^{\mu\nu\rho}
\eea 
Spinorial indices are lowered and raised as $(\g^\mu)^\a_{\; \, \b} = \varepsilon^{\a \g}  (\g^\mu)_\g^{\; \, \d} \varepsilon_{\b \d}$, where $\varepsilon^{12} = - \varepsilon_{12} = 1$. 
When writing spinorial products we conventionally choose the spinorial indices of chiral fermions to be always up, while the ones of antichirals to be always down.

\vskip 20pt
\noindent
The euclidean action  of $U(N)_k \times U(M)_{-k}$ ABJ(M) theory \cite{ABJM, ABJ} reads
 \bea
\label{action}
S &=& \frac{k}{4\pi}\int d^3x\,\varepsilon^{\mu\nu\rho} \Big\{ -i\Tr \left( A_\mu\partial_\nu A_\rho+\frac{2}{3} i A_\mu A_\nu A_\rho \right)
+i \Tr \left(\hat{A}_\mu\partial_\nu 
\hat{A}_\rho+\frac{2}{3} i \hat{A}_\mu \hat{A}_\nu \hat{A}_\rho \right) 
\non \\
&~& \qquad \qquad \qquad \qquad  + \Tr \Big[ \frac{1}{\xi}  (\pa_\mu A^\mu)^2 -\frac{1}{\xi} ( \pa_\mu \hat{A}^\mu )^2 + \pa_\mu \bar{c} D^\mu c  
  - \pa_\mu \bar{\hat{c}} D^\mu \hat{c} \Big] \Big\}
\non \\
&~& + \int d^3x \, \Tr \Big[ D_\mu C_I D^\mu \bar{C}^I + i \bar{\psi}^I \g^\mu D_\mu \psi_I \Big] + S_{int} 
\non  
\eea
with covariant derivatives defined as
\bea
\label{covariant}
D_\mu C_I &=& \pa_\mu C_I + i A_\mu C_I - i C_I \hat{A}_\mu
\quad ; \quad 
D_\mu \bar{C}^I = \pa_\mu \bar{C}^I - i \bar{C}^I A_\mu + i \hat{A}_\mu \bar{C}^I  
\non \\
D_\mu \bar{\psi}^I  &=& \pa_\mu \bar{\psi}^I + i A_\mu \bar{\psi}^I - i \bar{\psi}^I \hat{A}_\mu
\quad ; \quad
D_\mu \psi_I = \pa_\mu \psi_I - i \psi_I A_\mu + i \hat{A}_\mu \psi_I  
\eea
Gauge fields are the adjoint representation of the corresponding gauge group, $A_\mu = A_\mu^a T^a$, $\hat A_\mu = \hat A_\mu^a \hat T^a$ with $T^a$ ($\hat T^a$) a set of $U(N)$ ($U(M)$) hermitian matrices satisfying $\Tr (T^a T^b) = \d^{ab}$ ($\Tr (\hat T^a \hat T^b) = \d^{ab}$). Scalars $C_I$ ($\bar{C}^I$) and the corresponding fermions are in the (anti)bifundamental of the gauge group and carry a fundamental index of the $SU(4)$ R-symmetry group.  

With these assignments the Feynman rules are:
\begin{itemize}
\item Vector propagators in Landau gauge
\bea
\label{treevector}
&& \langle A_\mu^a (x) A_\nu^b(y) \rangle^{(0)} =  \d^{ab}   \, \left( \frac{2\pi i}{k} \right) \frac{\G(\frac32-\e)}{2\pi^{\frac32 -\e}} \varepsilon_{\mu\nu\rho} \frac{(x-y)^\rho}{[(x-y)^2]^{\frac32 -\e} }
\non \\
&& \langle \hat{A}_\mu^a (x) \hat{A}_\nu^b(y) \rangle^{(0)} =  -\d^{ab}   \, \left( \frac{2\pi i}{k} \right) \frac{\G(\frac32-\e)}{2\pi^{\frac32 -\e}} \varepsilon_{\mu\nu\rho} \frac{(x-y)^\rho}{[(x-y)^2]^{\frac32 -\e} }
\eea
\item Scalar propagator
\beq
\label{scalar}
\langle (C_I)_i^{\; \hat{j}} (x) (\bar{C}^J)_{\hat{k}}^l(\; y) \rangle^{(0)}  = \d_I^J \d_i^l \d_{\hat{k}}^{\hat{j}} \, \frac{\G(\frac12 -\e)}{4\pi^{\frac32-\e}} 
\, \frac{1}{[(x-y)^2]^{\frac12 -\e}}
\eeq
\item Fermion propagator
\beq
\label{treefermion}
\langle (\psi_I^\a)_{\hat{i}}^{\; j}  (x) (\bar{\psi}^J_\b )_k^{\; \hat{l}}(y) \rangle^{(0)} = - i \, \d_I^J \d_{\hat{i}}^{\hat{l}} \d_{k}^{j} \, 
\frac{\G(\frac32 - \e)}{2\pi^{\frac32 -\e}} \,  \frac{(\g^\mu)^\a_{\; \, \b} \,  (x-y)_\mu}{[(x-y)^2]^{\frac32 - \e}}
\eeq
\item Gauge cubic vertex
\beq
\label{gaugecubic}
-i \frac{k}{12\pi} \varepsilon^{\mu\nu\rho} \int d^3x \, f^{abc} A_\mu^a A_\nu^b A_\rho^c
\eeq
\item Gauge--fermion cubic vertex
\beq
\label{gaugefermion}
-\int d^3x \, \Tr \Big[ \bar{\psi}^I \g^\mu \psi_I A_\mu - \bar{\psi}^I \g^\mu \hat{A}_\mu \psi_I  \Big]
\eeq 
\end{itemize}

\noindent
For two--loop calculations we also need the one--loop vector propagators  
\bea
\label{1vector}
&& \langle A_\mu^a (x) A_\nu^b(y) \rangle^{(1)} = \d^{ab}   \left( \frac{2\pi }{k} \right)^2 N \frac{\G^2(\frac12-\e)}{4\pi^{3 -2\e}} 
\left[ \frac{\d_{\mu\nu}}{ [(x- y)^2]^{1-2\e}} - \pa_\mu \pa_\nu \frac{[(x-y)^2]^\e}{4\e(1+2\e)} \right]  \non  \\
&& \langle \hat{A}_\mu^a (x) \hat{A}_\nu^b(y) \rangle^{(1)} = \d^{ab}   \left( \frac{2\pi }{k} \right)^2 M \frac{\G^2(\frac12-\e)}{4\pi^{3 -2\e}} 
\left[ \frac{\d_{\mu\nu}}{ [(x- y)^2]^{1-2\e}} - \pa_\mu \pa_\nu \frac{[(x-y)^2]^\e}{4\e(1+2\e)} \right] \non \\
\eea
and the one--loop fermion propagator  
\beq
\label{1fermion}
\langle (\psi_I^\a)_{\hat{i}}^{\; j}  (x) (\bar{\psi}^J_\b)_k^{\; \hat{l}}(y) \rangle^{(1)} =  i \,\left( \frac{2\pi}{k} \right) \,  \d_I^J \d_{\hat{i}}^{\hat{l}} \d_{k}^{j} \,  \, \d^\a_{\; \, \b}
\, (N-M) \frac{\G^2(\frac12 - \e)}{16 \pi^{3-2\e}} \, \frac{1}{[(x-y)^2]^{1 - 2\e}}  
\eeq	 
 	 
\vskip 30pt	 
Given a generic (super)connection, the corresponding gauge invariant Wilson loop is 
\beq
\label{WL2}
W[\G] = \Str \Big[ \,P\, \exp{ \left( -i \int_\G d\tau {\cal L}(\tau)\right) } \, {\cal T} \Big]
\eeq 
where ${\cal T}$ is defined in \eqref{eq:T}, and for a latitude circle\footnote{Note that the path--ordering convention is opposite to the one used in \cite{DT,GMPS, Cardinali}.}
\beq
P\, \exp{ \left( -i \int_\G d\tau {\cal L}(\tau)\right) } \equiv 1 - i \int_0^{2\pi} d\tau \, {\cal L}(\tau) - \int_0^{2\pi} d\tau_1 \int_0^{\tau_1} d\tau_2  \, {\cal L}(\tau_1)  {\cal L}(\tau_2)  + \cdots
\eeq
We are interested in evaluating its vacuum expectation value
\beq
\langle  W[\G] \rangle \equiv \int  D[A, \hat{A}, C, \bar{C}, \psi, \bar{\psi}] \; e^{-S} \, W[\G]
\eeq

\section{Useful identities on the latitude circle}
\label{App:circle}

We parametrize a point on the latitude circle $\G$ as 
\beq
x_i^\mu = (\sin{\theta_0} , \cos{\theta_0} \cos{\tau_i},  \cos{\theta_0} \sin{\tau_i})  
\eeq 
Simple identities that turn out to be useful along the calculation are  
\begin{subequations}
\label{Iden1}
\bea
\label{I1}
&& (x_i - x_j)^2 = 4  \cos^2{\theta_0}  \, \sin^2{\frac{\tau_{ij}}{2}} =    ({\bf x}_i - {\bf x}_j)^2 \, \cos^2{\theta_0}
\\
\label{I2}
&& (x_i \cdot x_j) = \cos{\tau_{ij}} + \sin^2{\theta_0} \, (1 - \cos{\tau_{ij}})
\\
\label{I3}
&&  (\dot{x}_i \cdot \dot{x}_j) = \cos{\tau_{ij}} \, \cos^2{\theta_0} = (\dot{{\bf x}}_i \cdot \dot{{\bf x}}_j) \, \cos^2{\theta_0}
\\
\label{I4}
&& (x_i \cdot \dot{x}_j) = \sin{\tau_{ij}} \, \cos^2{\theta_0} = ({\bf x}_i \cdot \dot{{\bf x}}_j) \, \cos^2{\theta_0} 
\eea
\end{subequations}
where ${\bf x}_i^\mu = (0, \cos{\tau_i}, \sin{\tau_i})$ run on the maximal latitude ($\th=0$).

Using expression \eqref{eq:matrixfermionic} for the $\eta$ spinors, taking into account the following identities
\bea
\label{sidentity2}
s_I \bar{s}^I = -i
\quad , \quad (s_I \g^\mu \bar{s}^I) = 0 
\quad, \quad  (s_I \g^\mu \bar{s}^J) (s_J \bar{s}^I) = 0 
\quad, \quad   
(s_I \g^\mu \bar{s}^J) (s_J \g^\nu \bar{s}^I) = - \frac12 \d^{\mu\nu}
\non \\
\eea
and writing $\eta_i \equiv \eta(\tau_i)$, ${\mathcal M}_i  \equiv {\mathcal M}(\tau_i)$ a list of useful relations follows
\begin{subequations}
\label{Iden2}
\begin{align}
\label{id1}
& \eta_{i} \bar \eta_{j} = 
2 i\, \cos  \frac{\tau_{ij}}{2}\, e^{ i \nu \, \frac{\tau_{ij}}{2}} 
\left(\cos  \frac{\tau_{ij}}{2} - i\,  \nu\,  \sin \frac{\tau_{ij}}{2} \right)\\
& \eta_{i} \gamma_0 \bar \eta_{j} =  
-2i\, e^{ i \nu \frac{\tau_{ij}}{2} }    \, \sin  \frac{\tau_{ij}}{2}\, \left(
\nu\, \sin  \frac{\tau_{ij}}{2} + i\, \cos  \frac{\tau_{ij}}{2}\right)
\label{eq:etagamma0}\\
& \eta_{i} \gamma_1 \bar \eta_{j} =  \, e^{ i \nu \, \frac{\tau_{ij}}{2} }\left[ 
(\cos \t_i - \cos \t_j) \nu -  i  (\sin \t_i + \sin \t_j)
 \right]
\nonumber\\
& \eta_{i} \gamma_2 \bar \eta_{j} = i\,  \, e^{ i \nu \, \frac{\tau_{ij}}{2} } \left[ 
(\cos \t_i + \cos \t_j) - i\, \nu (\sin \t_i - \sin \t_j)
 \right]
\nonumber\\
& (\eta_{i} \gamma_{\mu} \bar \eta_{j})\, x_{ij}^{\mu} = 
2 i \,  \, \cos{\theta_0} \, \sin  \frac{\tau_{ij}}{2} 
\left[
(1-\nu) e^{ i (1+ \nu)\frac{\tau_{ij}}{2} }
+ (1+\nu) e^{ -i (1- \nu)\frac{\tau_{ij}}{2} }
\right] \nonumber\\
 &~~~~~ = 4 i\,  \, \cos{\theta_0} \, \sin  \frac{\tau_{ij}}{2}\,
\left[
\cos  \frac{\tau_{ij}}{2}\, \cos \left(  \nu\, \frac{\tau_{ij}}{2} \right)\,
+ \nu\, \sin \frac{\tau_{ij}}{2}\, \sin \left( \nu\, \frac{\tau_{ij}}{2} \right)
\right. \nonumber\\ & \left. ~~~~~~~
+i \left( 
\cos  \frac{\tau_{ij}}{2}\, \sin \left( \nu\, \frac{\tau_{ij}}{2} \right)
- \nu\, \sin \frac{\tau_{ij}}{2}\, \cos \left( \nu\, \frac{\tau_{ij}}{2}
 \right)\right)
\right]
 \label{eq:etagammax}
\\ 
\non \\
& \Tr({\cal M}_i {\cal M}_j) = 2 \left(1+\nu^2 + (1-\nu^2) \cos \tau_{ij} \right)
\label{matrices}
\end{align}
\end{subequations}
More generally, we can write
\bea
\label{eq:etagammaeta}
\left( \eta_i \gamma^{\mu} \bar\eta_j \right)
&=& i\, e^{i\nu\frac{\tau_{12}}{2}} \left( 1 - i\,\nu\, \tan \frac{\tau_{12}}{2} \right) \left( \frac{\dot x_1^{\mu}}{|\dot x_1|} + \frac{\dot x_2^{\mu}}{|\dot x_2|} + i\, \epsilon_{\lambda\nu\mu}\, \frac{\dot x_1^{\lambda}}{|\dot x_1|} \frac{\dot x_2^{\nu}}{|\dot x_2|}   \right)
\non \\
&=& i \frac{(\eta_i \g^0 \bar{\eta}_j)}{\sin (\tau_{ij})} \, \left( \frac{\dot x_i^{\mu}}{|\dot x_i|} + \frac{\dot x_j^{\mu}}{|\dot x_j|} + i\, \epsilon_{\lambda\nu\mu}\, \frac{\dot x_i^{\lambda}}{|\dot x_i|} \frac{\dot x_j^{\nu}}{|\dot x_j|}   \right)
\eea
In all these expressions the real and imaginary parts have definite (but different) parity under exchange $i\leftrightarrow j$.

\section{One--loop integrals}\label{App:oneloop}

Here we list the results for the one--loop integrals expanded up to finite terms in $\e$. Using the standard definitions
\beq
\label{eq:harmonics}
\Psi^{(n)}(z) = \frac{d^{n+1}}{dz^{n+1}} \log \G(z)  \qquad ; \qquad \Psi^{(0)}(1+x) = H_x - \gamma_E  
\eeq
where $H_x$ are the harmonic numbers, we can write
\bea
I_1 &=&
\frac{2 \pi  \nu-2 \sin (\pi  \nu)}{\epsilon } +
\frac{1}{\left(\nu^2\!-\!1\right)^2}\Bigg(\!-\!4\pi \nu \left(\nu^2\!-\!1\right) \left((\gamma\!-\!1) \nu^2+\left(\nu^2\!-\!1\right) \log (2)\!-\!\gamma\!+\!3\right)
 \nonumber\\&&
-\nu \left(\nu^2-1\right)^2 \sin (\pi  \nu) \left(\Psi ^{(1)}\left(-\frac{\nu}{2}-\frac{1}{2}\right)- \Psi ^{(1)}\left(\frac{\nu-1}{2}\right)\right)\nonumber\\&&+2 \left(\nu^2-1\right)^2 (\sin (\pi  \nu)-\pi  \nu) \Psi ^{(0)}\left(-\frac{\nu}{2}-\frac{1}{2}\right)
\nonumber\\& &\left.
+2 \left(\nu^2-1\right)^2 (\sin (\pi  \nu)-\pi  \nu) \Psi ^{(0)}\left(\frac{\nu-1}{2}\right)
\right.\nonumber\\&& +\left(-4 \left(\nu^4+3\right)+4 \gamma  \left(\nu^2-1\right)^2+4 \left(\nu^2-1\right)^2 \log (2)\right) \sin (\pi  \nu)\Bigg)
\\
\non \\
\non \\
I_2 
&=&
\frac{2 \pi }{\epsilon }-2 \pi  \left(H_{-\frac{\nu}{2}-\frac{1}{2}}+H_{\frac{\nu-1}{2}}+\log (4)\right)
+\sin (\pi  \nu) \left(\Psi ^{(1)}\left(\frac{\nu+1}{2}\right)-\Psi ^{(1)}\left(\frac{1}{2}-\frac{\nu}{2}\right)\right)
\non \\
\\
\non \\
I_3 
&=&
-\frac{2 (\cos (\pi  \nu)+1)}{\epsilon } +
\frac{1}{2 \left(\nu^2-1\right)^2}(\cos (\pi  \nu)+1)\times\nonumber\\&& \left(-8 \left(\nu^4+3\right)+8 \gamma  \left(\nu^2-1\right)^2+8 \left(\nu^2-1\right)^2 \log (2)+4 \left(\nu^2-1\right)^2 \Psi ^{(0)}\left(-\frac{\nu}{2}-\frac{1}{2}\right)
\right. \nonumber\\& &\left.
+2 \left(\nu^2-1\right)^2 \left(2 \Psi ^{(0)}\left(\frac{\nu-1}{2}\right)+\nu \left(\Psi ^{(1)}\left(\frac{\nu-1}{2}\right)-\Psi ^{(1)}\left(-\frac{\nu}{2}-\frac{1}{2}\right)\right)\right)\right)
\\
\non \\
\non \\
I_4 
&=&
(\cos (\pi  \nu)+1) \left(\Psi ^{(1)}\left(\frac{1}{2}-\frac{\nu}{2}\right)-\Psi ^{(1)}\left(\frac{\nu+1}{2}\right)\right)
\eea

\vskip 30pt  
\section{The fermionic two--loop diagrams}\label{app:fermionic}

In this appendix we spell out the computation of the diagrams entering the two--loop correction to the $1/6-$BPS fermionic Wilson loop \eqref{eq:fermionic}.
We proceed diagram by diagram expanding in detail all the relevant steps which led to the results presented in section \ref{sec:two-loop}.

\subsubsection*{One--loop fermion correction}
The simplest contribution originates from expanding the exponential of the superconnection at second order in the fermionic fields and contracting them with the one--loop corrected fermion propagator \eqref{1fermion} as depicted in Fig. \ref{fig:fermion}.  
\FIGURE[l]{
 \includegraphics[width=0.22\textwidth, height=3.8cm]{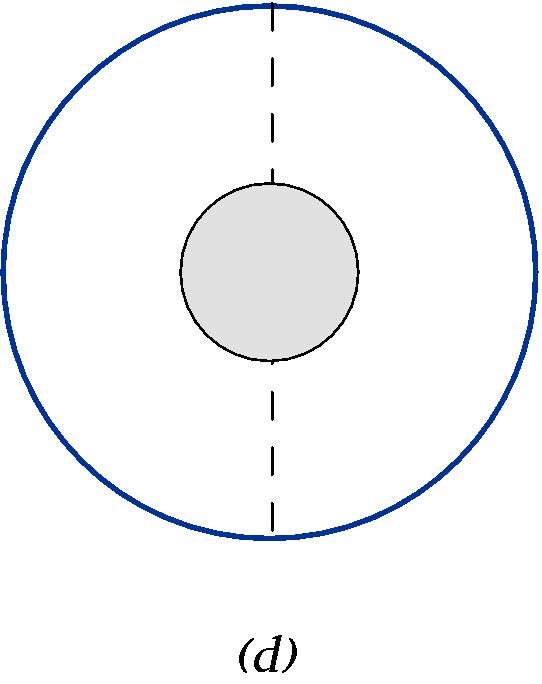}
    \caption{Exchange of a fermion with one--loop propagator.}
    \label{fig:fermion}
}
In the $1/2-$BPS case the exact cancellation between the contributions of the upper-left and the lower-right blocks of ${\cal L}(\tau_1) {\cal L}(\tau_2)$ leads to a vanishing result  \cite{BGLP3,GMPS}. For our deformed Wilson loop instead, such a mechanism no longer occurs because of the non--trivial dependence on the parameter $\nu$, which weights the combination of the two blocks.
Explicitly we have 
\begin{equation}
\begin{split}
{\rm (d)} =\frac{2\pi  {\cal R} }{k} \,  \int d\tau_{1>2}  {\rm tr} &\left[ \eta_{1 I} \bar{\eta}_2^J 
\langle  \bar{\psi}_1^I \psi_{2 J} \rangle^{(1)}  e^{-i\frac{\pi}{2} \nu}  -\right.\\ &\left. - \bar{\eta}_1^I \eta_{2 J} \langle \psi_{1\, I} \bar{\psi}_2^J \rangle^{(1)}  e^{i\frac{\pi}{2} \nu} \right] \, |\dot{x}_1| |\dot{x}_2|
\end{split}
\end{equation}

\noindent where $\int d\tau_{1>2} \equiv \int_0^{2\pi} d\tau_1 \int_0^{\tau_1} d\tau_2$.
Using expression \eqref{1fermion} for the one--loop fermion propagator and the identity \eqref{id1} for the $\eta$ spinors, we can write
\bea
\label{partiald}
{\rm (d)} &=&  {\cal R} \, (N-M) \, \frac{MN}{k^2} \, \frac{\G^2(\frac12 -\e)}{(4\p)^{1-2\e}} \, (\cos {\theta_0})^{4\e} \, \times
\\
&&\quad \Big\{  I_{(d)}[ 1 - 2\e,\nu]  -  \nu I_{(d)}[\tfrac12-2\e, \nu-1]  +(\nu-1)  I_{(d)}[- 2\e,\nu] \Big\}
\non
\eea
where we have defined
\beq
\label{integrald1}
I_{(d)}[\alpha,\nu]=\int d\tau_{1>2} \, 
\frac{\cos{ \frac{\nu (\tau_{12} -\pi)}{2} }}{\left(\sin^2 \frac{\tau_{12}}{2} \right)^{\alpha}}   
\eeq
The integral can be solved by expanding the trigonometric functions in power series \cite{BGLP3}. Performing the $\e$--expansion we obtain a finite result
\begin{equation}
{\rm (d)} =  {\cal R} \, (N-M) \, \frac{MN}{k^2} \, \pi \, \left( \nu - \frac{1}{\nu} \right) \sin{\frac{\pi \nu}{2}} 
\end{equation}

\subsubsection*{Double fermion exchange}
The forth order expansion of the Wilson loop exponential leads to two quartic terms 
\bea
&& {\cal R} \left( \frac{2\pi}{k} \right)^2 \int d\tau_{1>2>3>4} \, |\dot{x}_1| |\dot{x}_2| |\dot{x}_3| |\dot{x}_4|
\left[ \eta_{1 I} \bar{\eta}_2^J \eta_{3  K} \bar{\eta}_4^L \; \Tr \langle \bar{\psi}_1^I \psi_{2  J} \bar{\psi}_3^K \psi_{4 L} \rangle\, e^{-i \frac{\pi}{2} \nu} \right.
\non \\
&~& \left. \qquad \qquad \qquad \qquad - \,  \bar{\eta}_1^I \eta_{2 J} \bar{\eta}_3^K \eta_{4  L}  \; \Tr \langle  \psi_{1  I} \bar{\psi}_2^J \psi_{3 K} \bar{\psi}_4^L \rangle\, e^{i \frac{\pi}{2} \nu} \right]
\eea
where we can contract fermions in two possible ways, so obtaining the two diagrams in Fig. \ref{fig:fermion2}. 

\FIGURE{
 \includegraphics[width=0.34\textwidth]{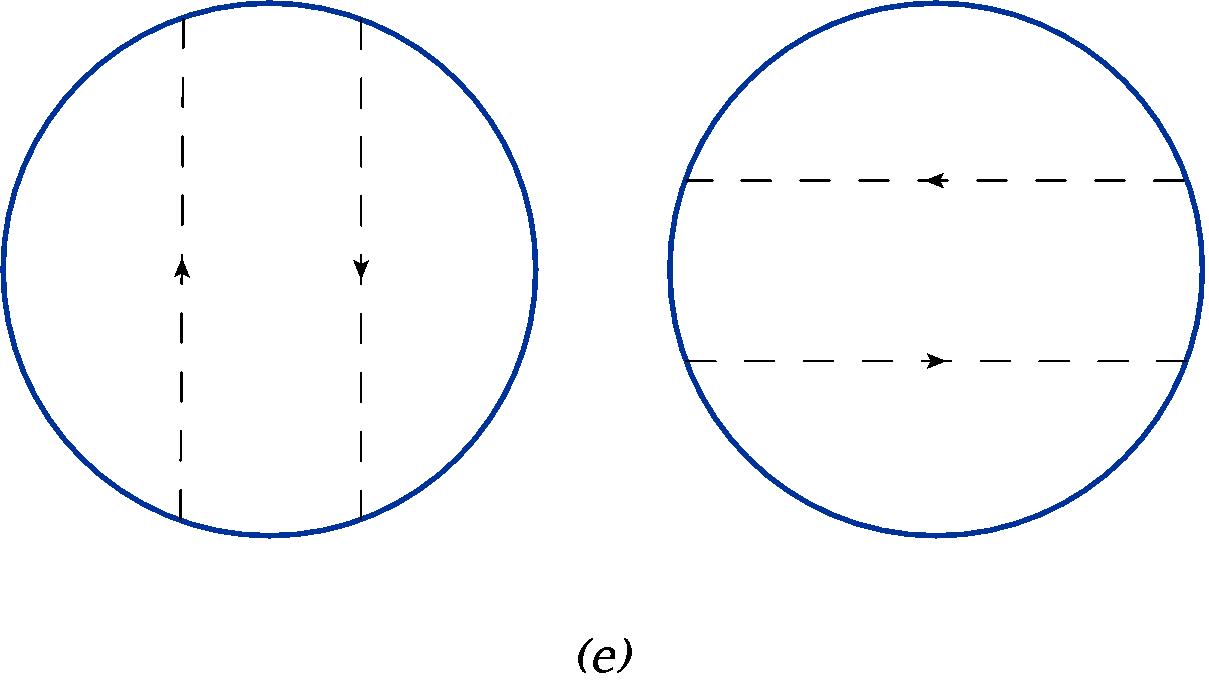}
    \caption{Two possible contractions for the double fermion exchange.}
    \label{fig:fermion2}
}
Their evaluation proceeds as for the $1/2-$BPS Wilson loop \cite{BGLP3, GMPS}, albeit the rather complicated form of the spinors $\eta$ triggers a nasty proliferation of terms with different four-fold integrals in the loop parameters.

In order to perform the computation in a compact way, we find convenient to express the fermion propagator as 
\beq
\langle (\psi_I^\a)_{\hat{i}}^{\; j}  (x) (\bar{\psi}^J_\b )_k^{\; \hat{l}}(y) \rangle^{(0)} =  i \, \d_I^J \d_{\hat{i}}^{\hat{l}} \d_{k}^{j} \, 
(\g^\mu)^\a_{\; \, \b} \,  \pa_\mu D(x-y)
\eeq
where
\beq
D(x) =  \frac{\G(\frac12 - \e)}{4\pi^{\frac32 -\e}} \frac{1}{(x^2)^{\frac12 - \e}}
\eeq
and to rearrange the  {\it ubiquitous} factor $(\eta_i \gamma^{\mu} \bar\eta_j) \, \partial_{\mu}\, D(x_{ij})$  as follows
\begin{equation}
\cos^2 \theta_0\, (\eta_i \gamma^{\mu} \bar\eta_j) \, \partial_{\mu}\, D(x_{ij}) = \frac{d}{d\tau_i} g_{\epsilon}(\tau_{ij}) - i\, \nu \epsilon g_{\epsilon}(\tau_{ij})
\end{equation}
The function
\begin{equation}
g_{\epsilon}(\tau) = i\,  l (\cos \theta_0)^{2\epsilon}  \frac{\Gamma\left(\tfrac12-\epsilon\right) e^{\frac{i \nu \tau}{2}}}{4^{1-\epsilon} \pi^{\tfrac32-\epsilon} \left( \sin^2 \frac{\tau}{2} \right)^{\frac12-\epsilon}}
\end{equation}
satisfies the reflection property $g_{\epsilon}(-\tau) = -\overline{g_{\epsilon}(\tau)}$.

The contribution from diagram (e) can then be rewritten as  
\begin{align}
& {\rm (e)} = - {\cal R} \left( \frac{2\pi}{k} \right)^2  \, \int d\tau_{1>2>3>4} \, \bigg\{  
e^{-i \nu \frac{\pi}{2}} \left[ 
NM^2 f_{\epsilon}(\tau_1,\tau_3) - M N^2 f_{\epsilon}(\tau_3,\tau_1) \right] 
\nonumber\\
& ~~~~~~~~~~~~~~~~~~~~~~~~\qquad \qquad
- e^{i \nu \frac{\pi}{2}} \left[ 
MN^2 \overline{f_{\epsilon}}(\tau_1,\tau_3) - N M^2 \overline{f_{\epsilon}}(\tau_3,\tau_1) \right]  \bigg\}
\end{align}
where
\begin{equation}
f_{\epsilon}(\tau_1,\tau_3) = \left( \frac{d}{d\tau_1} g_{\epsilon}(\tau_{12}) - i\, \nu \epsilon g_{\epsilon}(\tau_{12}) \right) \left( \frac{d}{d\tau_3} g_{\epsilon}(\tau_{34}) - i\, \nu \epsilon g_{\epsilon}(\tau_{34}) \right)
\end{equation}
By performing the products of the two factors this expression gives rise to eight four-fold integrals, which can be nevertheless all related to each other. 
We choose as reference integral
\beq
I_{(e)} \equiv \int_{0}^{2\pi} d\tau_1 \int_{0}^{\tau_1} d\tau_2 \int_{0}^{\tau_2} d\tau_3 \int_{0}^{\tau_3} d\tau_4 \frac{d}{d\tau_1} g_{\epsilon}(\tau_{12}) \frac{d}{d\tau_3} g_{\epsilon}(\tau_{34})
\eeq
and express all the others in terms of it. First of all, we perform two easy integrations and rewrite
\begin{equation}
I_{(e)} = \int_{0}^{2\pi} d\tau_1 \int_{0}^{\tau_1} d\tau_2\, \frac{e^{i \nu \frac{\tau_2-\tau_1}{2}}}{\sin^{1-2\epsilon}{\frac{\tau_2}{2}} \sin^{1-2\epsilon}{\frac{\tau_1}{2}}}\, e^{i \nu \pi}
\end{equation}
Proceeding similarly for the other seven integrals and playing with suitable change of integration variables allows to rewrite the contribution of diagram (e) as
\begin{align}
\label{intermidiate(e)}
& {\rm (e)} =  - {\cal R} \left( \frac{2\pi}{k} \right)^2  \, \frac{\G^2(\frac12 -\e)}{4^{2-2\e} \pi^{3-2\e}} \, (\cos \theta_0)^{4\e} \, 2\pi i \nu M N \times
\\
& \left\{ M \left[ \epsilon\, e^{-i\nu\tfrac{\pi}{2}} I_{(e)}+\epsilon^2\, \nu \,  e^{i\nu\tfrac{\pi}{2}} 
\frac{d}{d\nu}(e^{-i\nu\pi}I_{(e)}) \right]\!
 +\! N \left[ \epsilon\, e^{i\nu\tfrac{\pi}{2}} I_{(e)}^* +\epsilon^2\, \nu \,  e^{-i\nu\tfrac{\pi}{2}} 
\frac{d}{d\nu}(e^{i\nu\pi}I_{(e)}^*) \right] \right\} \non
\end{align}
The resulting two-fold integral can be solved in terms of hypergeometric series, following the techniques of \cite{BGLP3}. The procedure is quite long but straightforward and leads to the following exact result
\begin{align}
I_{(e)}  &= 
i\pi e^{i\pi\nu} 2^{3-4 \epsilon } \left(\frac{ 
_3F_2\left(
\begin{array}{c}
1\!-\!2 \epsilon ,1\!-\!2 \epsilon ,\!-\!\epsilon\!-\!\frac{\nu }{2}\!+\!\frac{1}{2} \\
1,\!-\!\epsilon\!-\!\frac{\nu }{2}\!+\!\frac{3}{2}
\end{array} \right)}{-\nu -2 \epsilon +1}
-\frac{ 
_3F_2\left(
\begin{array}{c}
1\!-\!2 \epsilon ,1\!-\!2 \epsilon ,\!-\!\epsilon\!+\!\frac{\nu }{2}\!+\!\frac{1}{2}\\
1,\!-\!\epsilon\!+\!\frac{\nu }{2}\!+\!\frac{3}{2}
\end{array} \right)}{\nu -2 \epsilon +1}
\right) 
\nonumber\\&
+2^{2-4\epsilon}\Gamma^2(2\epsilon)(e^{i\pi\nu}\cos(2\pi\epsilon)+1)
\frac{\Gamma \left(-\epsilon +\frac{\nu }{2}+\frac{1}{2}\right)
\Gamma \left(-\epsilon -\frac{\nu }{2}+\frac{1}{2}\right)}
{\Gamma \left(\epsilon +\frac{\nu }{2}+\frac{1}{2}\right)
\Gamma \left(\epsilon -\frac{\nu }{2}+\frac{1}{2}\right)}
\end{align}
After analytic continuation of the hypergeometric series, we can expand $I_{(e)}$ around $\epsilon=0$. Keeping terms up to finite orders, from eq. \eqref{intermidiate(e)} we obtain the final expression for diagram (e) 
\begin{align}
{\rm (e)} &= -{\cal R}  \,NM(N + M)\, \left(\frac{2\pi}{k}\right)^2   
\frac{\Gamma^2(\tfrac{1}{2}-\epsilon)}{16 \pi^{3-2\epsilon}}   \, 
\frac{4i\pi\nu\cos(\tfrac{\pi\nu}{2})}{\epsilon} \,  (\cos \theta_0)^{4\e}
\nonumber\\
&~~~+{\cal R}  \, MN (N + M) \frac{i\pi\nu}{2 k^2} 
\left[ \pi(\nu-4)\sin(\tfrac{\pi\nu}{2})+8\cos(\tfrac{\pi\nu}{2})
\,  H_{\tfrac{\nu}{2} -\tfrac12} \right]
\non \\
&~~~- {\cal R} NM(M - N) \frac{\pi^2 \nu^2}{2k^2} \, \cos{\frac{\pi\nu}{2}}
\end{align}
where $H_n$ are the harmonic numbers (see eq. \eqref{eq:harmonics}). 
Contrary to the $1/2-$BPS case, this contribution is divergent, as signalled by the $\epsilon$--pole which consistently disappears in the $\nu\rightarrow 1$ limit.

\bigskip

\bigskip

\subsubsection*{Vertex diagram}

The most involved part of the perturbative evaluation of our fermionic Wilson loop comes from the vertex diagram of Figure \ref{fig:vertex}.
Considering all terms coming from the cubic expansion of the operator this diagram corresponds to the following contributions
\begin{align}
\label{3expansion}
& - i {\cal R}  \left( \frac{2\pi}{k} \right) \, \cos^2{\theta_0} \, \int d\tau_{1>2>3} \, \times
\\  
& {\rm tr} 
\Big\{  e^{-i\nu \frac{\pi}{2}} \left[ \eta_{2  I} \bar{\eta}_3^J  \, \langle A_{1\mu} \bar{\psi}_2^I \psi_{3  J} \rangle \, \dot{x}_1^\mu  
~+~ \bar{\eta}_3^I \eta_{1 J}  \, \langle \bar{\psi}_1^J \hat{A}_{2 \mu} \psi_{3 I}  \rangle \, \dot{x}_2^\mu  
~+~ \eta_{1 I} \bar{\eta}_2^J  \, \langle  \bar{\psi}_1^I \psi_{2 J} A_{3 \mu} \rangle \, \dot{x}_3^\mu  \right]
\non \\
& ~-~  e^{i\nu \frac{\pi}{2}} \left[  \bar{\eta}_2^I \eta_{3 J}    \, \langle   \hat{A}_{1 \mu}  \psi_{2 I} \bar{\psi}_3^J    \rangle \, \dot{x}_1^\mu   
~+~ \eta_{3 I}  \bar{\eta}_1^J   \, \langle  \psi_{1 I} A_{2\mu}   \bar{\psi}_3^J   \rangle \, \dot{x}_2^\mu  
~+~ \bar{\eta}_1^I \eta_{2 J}  \, \langle \psi_{1  I} \bar{\psi}_2^J \hat{A}_{3\mu} \rangle \, \dot{x}_3^\mu \right] \,  \Big\} 
\non
\end{align}
\FIGURE{
 \includegraphics[width=0.2\textwidth]{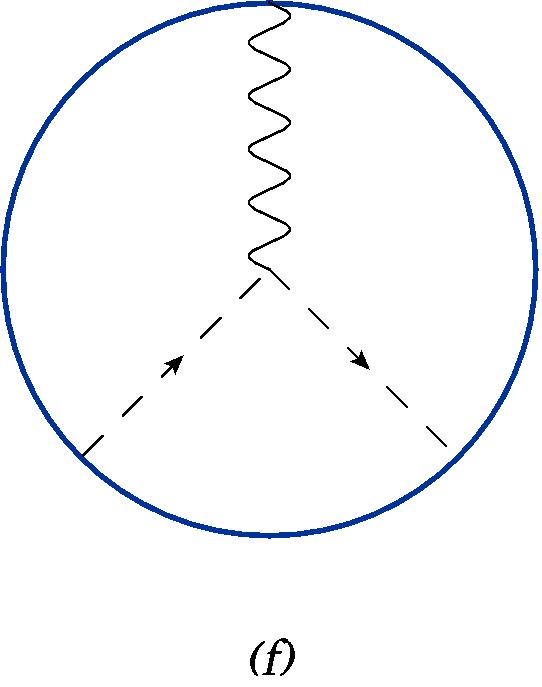}
    \caption{Vertex diagram.}
    \label{fig:vertex}
}
where the expectation value entails a contraction with the cubic interaction vertex \eqref{gaugefermion}.
Evaluating it explicitly as in \cite{BGLP3,GMPS} we end up with an expression which is proportional to spinorial structures of the form $(\eta_i \g_\l \g^\mu \g_\nu \bar \eta_j)$. Exploiting their symmetry under exchange of $i \leftrightarrow j$ as follows from identity (\ref{beginningf}), it is easy to realize that the contributions proportional to $NM^2$ can be easily obtained 
from the ones proportional to $MN^2$ by sending $\nu \rightarrow -\nu$ and multiplying by an overall minus sign.\\ Therefore, we can concentrate only on the $MN^2$ terms 
\begin{align}
\label{beginningf}
& {\cal R}\, MN^2 \left( \frac{2\pi}{k} \right)^2 \cos^2{\theta_0}  \, \int d\tau_{1>2>3} \, \times 
\Big[ \left( \eta_1 \gamma_\lambda\gamma^\mu\gamma_\nu \bar\eta_2 \right) \, \epsilon_{\mu\rho\sigma} \, \dot x_3^\rho\,  \Gamma^{\lambda\nu\sigma}\, e^{-i\frac{\pi\nu}{2}}+ 
\nonumber\\&
+ 
\left( \eta_2 \gamma_\lambda\gamma^\mu\gamma_\nu \bar\eta_1 \right) \, \epsilon_{\mu\rho\sigma} \, \dot x_1^\rho\, \Gamma^{\sigma\lambda\nu} \, e^{-i\frac{\pi\nu}{2}}
+ \left( \eta_3 \gamma_\lambda\gamma^\mu\gamma_\nu \bar\eta_1 \right) \, \epsilon_{\mu\rho\sigma} \, \dot x_2^\rho\,  \Gamma^{\nu\sigma\lambda}\, e^{i\frac{\pi\nu}{2}} \Big] 
\end{align}
where we have defined
\begin{equation}\label{gammaintegrals}
\Gamma^{\mu\nu\rho} \equiv \,
\left(\frac{\Gamma\left( \tfrac12 - \epsilon \right)}{4\pi^{\tfrac32-\epsilon}}\right)^3\, \partial^{\mu}_1\, \partial^{\nu}_2\, \partial^{\rho}_3\, \int \frac{d^3 x}{[ (x-x_1)^2 (x-x_2)^2 (x-x_3)^2]^{\frac12 -\e}} 
\end{equation}
Concentrating for instance on the first term in \eqref{beginningf} (the other two terms are obtained by simply permuting the indices) we observe that 
\begin{align}
\label{gammaid}
&\left( \eta_1 \gamma_\lambda\gamma^\mu\gamma_\nu \bar\eta_2 \right) \, \epsilon_{\mu\rho\sigma} \, \dot x_3^\rho \, \Gamma^{\lambda\nu\sigma} =
\\&
= - \left( \eta_1 \gamma^\mu \bar\eta_2 \right) \epsilon_{\mu\rho\sigma} \dot x_3^\rho \Gamma^{\nu\phantom{\nu}\sigma}_{\phantom{\nu}\nu}
- i \left( \eta_1 \bar\eta_2 \right)\, \dot x_3^\rho \left( \Gamma^\lambda_{\phantom{\lambda}\rho\lambda}
- \Gamma_{\rho\phantom{\lambda}\lambda}^{\phantom{\rho}\lambda} \right)
+ \left( \eta_1 \gamma_\nu \bar\eta_2 \right) \epsilon_{\mu\rho\sigma} \dot x_3^\rho \left( \Gamma^{\mu\nu\sigma} + \Gamma^{\nu\mu\sigma}  \right)
\non \\&
= \underbrace{ \left( \eta_1 \gamma^0 \bar\eta_2 \right) x_{3 \,\s} \Gamma^{\nu\phantom{\nu}\s}_{\phantom{\nu}\nu}}_{A)} 
\underbrace{- i \left( \eta_1 \bar\eta_2 \right)\, \dot x_3^\rho \left( \Gamma^\lambda_{\phantom{\lambda}\rho\lambda}
- \Gamma_{\rho\phantom{\lambda}\lambda}^{\phantom{\rho}\lambda} \right)}_{B)} 
+\underbrace{\left( \eta_1 \gamma_\nu \bar\eta_2 \right) \epsilon_{\mu\rho\sigma} \dot x_3^\rho \left( \Gamma^{\mu\nu\sigma} + \Gamma^{\nu\mu\sigma}  \right)}_{C)} 
\non
\end{align}
where in the last line we have used the planarity of the path and the identities of appendix B to simplify the first term. 

Whenever a $\Gamma$ integral has two contracted indices, as in the pieces A) and B), its computation simplifies considerably, since the integral can be solved using the Green equation obtaining \cite{Griguolo}
\beq
\Gamma^{\nu\mu}_{\phantom{\nu\mu}\mu} =  \partial_1^{\nu} \, \Phi_{1;23} \qquad  \qquad 
\Gamma^{\mu}_{\phantom{\mu}\nu\mu} =  \partial_2^{\nu} \, \Phi_{2;13} \qquad  \qquad 
\Gamma^{\mu}_{\phantom{\mu}\mu\nu} =  \partial_3^{\nu} \, \Phi_{3;12}
\eeq
where
\beq
\Phi_{i;jk} =  - \frac{\G^2(\tfrac12 -\e)}{32 \pi^{3 -2\e}}  \, \left[ \frac{1}{(x_{ij}^2 x_{ik}^2)^{1/2 -\e}}   - \frac{1}{(x_{ij}^2 x_{jk}^2)^{1/2 -\e}}  -\frac{1}{(x_{ik}^2 x_{jk}^2)^{1/2 -\e}}  \right]
\eeq
It is convenient to discuss separately the contributions A), B) or C) in \eqref{gammaid}.

The contracted contributions A) and B) are simpler to evaluate, since they do not feature a space-time integration any longer.  
We have to compute triple $\tau$--integrals of linear combinations of $\Phi_{i;jk}$ functions multiplied by spinorial structures. Using identities \eqref{I1} and (\ref{id1}--\ref{eq:etagammax}), the integrands can be expressed as products of trigonometric functions. Due to the involved form of the spinors $\eta$ their number is considerably higher than the $1/2$--BPS case.
 We evaluate these integrals by using the prescription of \cite{BGLP3, GMPS}. In order to simplify the computation we apply the following strategy: First we identify terms which can be expressed as total $\tau$--derivatives and perform integrations trivially. In the rest of the pieces, we exploit symmetries of the integrands to easily perform some integration. This turns out to be always possible and gives double integrals at most.
Finally we observe that, after such a reduction, non--trivial partial cancellations occur when summing the three different permutations in \eqref{beginningf} and the resulting integrals are in general simpler than the individual ones. 

Such integrals display non--trivial numerators. Applying ordinary trigonometric identities they can be reduced to the sum of contributions where denominators get cancelled by analogous expressions at numerators plus some remaining. When denominators are cancelled we obtain manifestly finite integrals, so we can evaluate them straightforwardly at $\epsilon=0$.
The remaining integrals which require a solution for generic $\epsilon$ are faced with the technique of \cite{BGLP3}. Namely, we first expand trigonometric functions in power series; then, sitting in safe regions of the $\e$--plane, we compute the integrals term by term. The results can be summed in terms of hypergeometric functions. Using their properties we can perform analytic continuation in such a way that they converge in a region around $\epsilon=0$. In such a form they can be safely expanded and the expansion truncated at finite order in the dimensional regularization parameter.

The intermediate steps of this procedure turn out to be rather lengthy but straightforward, so we skip the details providing only the final results.

The total $\epsilon$ expansion of the A) contracted integral reads  
\begin{align}
\label{firstcontracted}
A) & = i\, {\cal R}\, M N^2\, \left(\frac{2\pi}{k}\right)^2  \frac{\Gamma^2( \tfrac12-\epsilon)}{32 \pi^{3-2\epsilon}} \, \left\{ 
\frac{8 \pi  \nu  \left(\nu ^2-3\right) \cos \left(\frac{\pi  \nu }{2}\right)}{\nu ^2-1} \left(\frac{1}{\epsilon } -4 H_{-\frac{\nu+1}{2}} -\right.\right.\nonumber\\ &\left.\left.-2 \pi  \tan \left(\frac{\pi  \nu }{2}\right)\right)
\frac{32 \pi  \nu  \left(\nu ^2-5\right) \cos \left(\frac{\pi  \nu }{2}\right)}{\left(\nu ^2-1\right)^2} - 8 i \pi  \left(\frac{1}{\nu }-\nu \right) \sin \left(\frac{\pi  \nu }{2}\right)
\right\}
\end{align}
whereas the expansion for the B) integral gives 
\begin{align}
\label{secondcontracted}
B) & = {\cal R}\, MN^2 \, \left(\frac{2\pi}{k}\right)^2 \, \facc  
\bigg\{
\frac{8 i \pi  \nu}{\nu ^2-1}
\left[ \frac{\cos \left(\frac{\pi  \nu }{2}\right)}{\epsilon } 
\right.\nonumber\\& \left.
~~~~ -2 \cos \left(\frac{\pi  \nu }{2}\right) \left(3 H_{-\frac{\nu }{2}-\frac{1}{2}}-\frac{i \pi}{2} +2 \log 2\right) + 2\, i \sin \left(\frac{\pi  \nu }{2}\right) \left(H_{\frac{\nu }{2}-1}+2 \log 2\right)\right] 
\nonumber\\&
~~~~ +\frac{8 \pi  \left[(\nu -1)^2-3 i \pi  \nu ^2\right] \sin \left(\frac{\pi  \nu }{2}\right)}{\nu  \left(\nu ^2-1\right)}+\frac{24 i \pi  \nu  \left(\nu ^2+3\right) \cos \left(\frac{\pi  \nu }{2}\right)}{\left(\nu ^2-1\right)^2}
\bigg\}
\end{align}
A non--trivial consistency check of these expressions is provided by taking the $\nu\rightarrow 1$ limit. It is easy to see that in this limit the results collapse to the ones for the 
$1/2-$BPS Wilson loop \cite{BGLP3, GMPS}.

\bigskip

Finally, we have to evaluate 
\begin{align}
\label{uncontracted}
C) + {\rm perm.} & = {\cal R}\, \left(\frac{2\pi}{k}\right)^2 MN^2  \left[ \left( \eta_1 \gamma_{\nu} \bar\eta_2 \right)\, \varepsilon_{\mu\rho\sigma} \dot x_3^\rho \left( \Gamma^{\mu\nu\sigma} + \Gamma^{\nu\mu\sigma} \right) e^{-i \frac{\pi\nu}{2}}
+ \right.\\& \left.
\left( \eta_2 \gamma_{\nu} \bar\eta_3 \right)\, \varepsilon_{\mu\rho\sigma} \dot x_1^\rho \left( \Gamma^{\sigma\mu\nu} + \Gamma^{\sigma\nu\mu} \right) e^{-i \frac{\pi\nu}{2}} + 
\left( \eta_3 \gamma_{\nu} \bar\eta_1 \right)\, \varepsilon_{\mu\rho\sigma} \dot x_2^\rho \left( \Gamma^{\nu\sigma\mu} + \Gamma^{\mu\sigma\nu} \right) e^{i \frac{\pi\nu}{2}}
\right]
\non 
\end{align}
This contribution is the hardest one since we have to solve a space-time integral, first. 
Such an integral is divergent and hence we have in principle to $\epsilon$--regularize it.

A convenient approach to evaluate this contribution was derived in \cite{GMPS} for the $1/2-$BPS case.
The strategy consists in adding and subtracting a suitable divergent integrand that is easier to evaluate in dimensional regularization and that, once subtracted, renders our space-time integral finite (in the sense that all $x_{ij}^2\rightarrow 0$ limits are not singular).
As stressed in \cite{GMPS} this subtraction has to regularize the coincident points limits at all orders in $\epsilon$, in order not to neglect evanescent terms.

Following \cite{GMPS}, we manipulate expression (\ref{uncontracted}) by using identity \eqref{eq:etagammaeta}. Because of the planarity of the path, only the last term of that identity contributes. 
Taking for instance the first term appearing in \eqref{uncontracted} we can write 
\beq
\label{integral}
 \left( \eta_1 \gamma_{\nu} \bar\eta_2 \right)\, \varepsilon_{\mu\rho\sigma} \dot x_3^\rho \left( \Gamma^{\mu\nu\sigma} + \Gamma^{\nu\mu\sigma} \right) =   
\frac12 (\eta_1 \g^0 \bar{\eta}_2) \, \left[ (x_{23}^2 + x_{13}^2 ) h_{\mu} h_{\nu} \mathbf{V}^{\mu\nu} \right]
\eeq
where $h^{\mu} = \delta_3^{\mu}$ and
\begin{align}
\label{Vmn}
\mathbf{V}^{\mu\nu}
= -\left(\frac{\Gamma(\frac{3}{2}-\epsilon)}{2\pi^{3/2-\epsilon}}\right)^{3}
\int d^{3-2\epsilon}w
\frac{w^{\mu}w^{\nu}}{(x_{1w}^{2})^{3/2-\epsilon}(x_{2w}^{2})^{3/2-\epsilon}(x_{3w}^{2})^{3/2-\epsilon}}
\end{align}
At this point, the main observation is that  the expression
\beq
F_{12,3} \equiv 
\frac12 (\eta_1 \g^0 \bar{\eta}_2) \, 
\left[ (x_{23}^2 + x_{13}^2 ) h_{\mu} h_{\nu} \mathbf{V}^{\mu\nu} -2 (1-2\epsilon)\, x_{3\mu} \Gamma^{\rho\phantom{\rho}\mu}_{\phantom{\rho}\rho} \right]
\eeq
is completely finite at coincident points, for any value of $\epsilon$.
Therefore, adding and subtracting the second term to the integral (\ref{integral}) we can write
\beq
\left( \eta_1 \gamma_{\nu} \bar\eta_2 \right)\, \varepsilon_{\mu\rho\sigma} \dot x_3^\rho \left( \Gamma^{\mu\nu\sigma} + \Gamma^{\nu\mu\sigma} \right) =
F_{12,3} 
+  (1-2\epsilon)\, (\eta_1 \g^0 \bar{\eta}_2) \,   x_{3\mu} \Gamma^{\rho\phantom{\rho}\mu}_{\phantom{\rho}\rho} 
\eeq
The extra piece has exactly the same form of the contracted integral A) in eq. \eqref{gammaid}. Therefore, this addition simply requires multiplying the result \eqref{firstcontracted} by an extra factor $1+ (1-2\e) = 2(1-\e)$.

As a final step we are left with the evaluation of the finite contribution
\begin{equation}
C) + {\rm perm.} \rightarrow {\cal R}\, \left(\frac{2\pi}{k}\right)^2 MN^2 \, 
\int d\tau_{1>2>3} \left[ e^{-i\,\nu\,\frac{\pi}{2}} \left( F_{12,3} + F_{23,1} \right) - e^{i\,\nu\,\frac{\pi}{2}} F_{31,2} \right]
\end{equation}
for which we find 
\begin{equation}
h_{\mu} h_{\nu} \mathbf{V}^{\mu\nu} = 
\frac{1}{16 \pi ^2 \sqrt{x_{12}^{2}}\sqrt{x_{13}^{2}}\sqrt{x_{23}^{2}}
   \left(\sqrt{x_{12}^{2}}+\sqrt{x_{13}^{2}}+\sqrt{x_{23}^{2}}\right)}
\end{equation}
We can perform the $\tau$--integrations using techniques similar to the ones used for the contracted terms. The explicit evaluation requires some work, but eventually it leads to a rather simple result
\begin{align}
\label{eq:uncontracted}
C) & =
- \frac{ i\, \pi MN^2{\cal R} }{k^2}\, 
\bigg\{
\frac{1}{\nu  \left(\nu ^2\!-\!1\right)^2}  \left[4 \nu ^2 \left(\nu ^2\!+\!1\right) \cos \left(\frac{\pi  \nu }{2}\right)\!+\!2 i \left(1\!-\!\nu ^2\right) (\nu \!-\!1)^2 \sin \left(\frac{\pi  \nu }{2}\right)\right]  
\nonumber\\&
+\frac{2\nu}{\nu ^2-1}  \left[ i \sin \left(\frac{\pi  \nu }{2}\right) \left(2 H_{\frac{\nu }{2}-1}\!+\!i \pi \!+\!4 \log 2\right)\!-\! \cos \left(\frac{\pi  \nu }{2}\right) \left(2 H_{-\frac{\nu+1}{2}}\!-\!i \pi \!+\!4 \log 2\right)\right]\!\!
\bigg\}
\end{align} 

\bigskip

We can now derive the complete expression for diagram (f) by combining the different pieces. In order to obtain the terms proportional to $MN^2$ 
we multiply the contribution \eqref{secondcontracted} by the extra factor $2(1-\epsilon)$ and sum it to \eqref{firstcontracted} and to the finite terms \eqref{eq:uncontracted}.
The contributions proportional to $NM^2$ can be easily obtained from this result by exchanging $M \leftrightarrow N$, $\nu \to -\nu$ and putting an overall minus sign. 

Summing everything we finally obtain
\begin{align}
{\rm (f)}&= {\cal R}\, MN (N + M)  \left(\frac{2\pi}{k}\right)^2\, \frac{\Gamma^2( \tfrac12-\epsilon)}{16 \pi^{3-2\epsilon}}  \, \frac{4 i \pi \nu \cos \left(\frac{\pi  \nu }{2}\right)}{\e} \, (\cos{\th})^{4\e} 
\\
&~ + {\cal R} \, MN(N+ M) \, \frac{i\pi\nu}{2k^2} \left[  4\pi \sin \left(\frac{\pi  \nu }{2}\right) - 8\cos \left(\frac{\pi  \nu }{2}\right) \,  H_{\tfrac{\nu}{2} -\tfrac12} \right]
\nonumber \\
&~ - {\cal R} \, \frac{MN(N -  M)}{k^2} \, \frac{\pi(\nu^2-1)}{\nu} \sin \left(\frac{\pi  \nu }{2}\right).
\nonumber
\end{align}

\section{Weak coupling expansions}\label{sec:expansions}

In this appendix we provide formulae for the weak coupling expansion of the $n$-wound $1/6-$BPS Wilson loop and the Bremsstrahlung function for the $1/2-$BPS cusp, obtained from our conjecture \eqref{crazy}.

From the localization results of \cite{MarinoPutrov} expanded at $\lambda \ll 1$, we find
\begingroup
\allowdisplaybreaks
\begin{align}
& \langle W_n^{1/6} \rangle = 1 + i \pi   m^2  \lambda  +  \left(\frac{2 \pi ^2 m^2}{3}  -\frac{\pi ^2 m^4}{3}\right) \lambda ^2  -\frac{1}{18} i \pi ^3   m^2 \left(m^4-8 m^2+4\right) \lambda ^3+
\nonumber\\&
+\frac{\pi ^4\lambda ^4}{180} m^2 \left(m^6-20 m^4+58 m^2-60\right) + 
\nonumber \\&
+\frac{i \pi ^5 \lambda ^5}{2700} m^2 \left(m^8-40 m^6+328 m^4-960 m^2+566\right)  + 
\nonumber\\&
-\frac{\pi ^6 \lambda ^6}{56700} m^2 \left(m^{10}-70 m^8+1218 m^6-8080 m^4+20566 m^2-21420\right)  +
\nonumber\\&
-\frac{i \pi ^7 m^2 \lambda ^7}{1587600} \left(m^{12}-112 m^{10}+3528 m^8-45376 m^6+269416 m^4-743232 m^2+461280\right)
\nonumber\\&
+\frac{\pi ^8 \lambda ^8}{57153600} m^2 \left(m^{14}-168 m^{12}+8652 m^{10}-192824 m^8+2156658 m^6-12496008 m^4 +
\right.\nonumber\\&\left.~~~~
+31198084 m^2-32621400\right) + 
\nonumber\\&
+\frac{i \pi ^9 \lambda ^9}{2571912000} m^2 \left(m^{16}-240 m^{14}+18816 m^{12}-669440 m^{10}+12569064 m^8 +
\right.\nonumber\\&\left.~~~~
-131449920 m^6+725469760 m^4-1963814400 m^2+1250353314\right) +
\nonumber\\&
-\frac{\pi ^{10} \lambda ^{10}}{141455160000} m^2 \left(m^{18}-330 m^{16}+37356 m^{14}-1994960 m^{12}+58274106 m^{10} +
\right.\nonumber\\&\left.~~~~
-995215740 m^8+9886401316 m^6-54619693920 m^4+135520273746 m^2 +
\right.\nonumber\\&\left.~~~~
-142086968100\right) + 
\nonumber\\&
+\frac{i \pi ^{11} \lambda ^{11}}{42436548000} m^2 \left(55 m^{16}-13200 m^{14}+1214268 m^{12} +
\right.\nonumber\\&\left.~~~~
-56910656 m^{10}+1490748864 m^8-22032434688 m^6+175649551363 m^4 +
\right.\nonumber\\&\left.~~~~
-663514816536 m^2+949696112700\right) + {\cal O}\left(\lambda ^{12}\right)  
\end{align}
\endgroup
Plugging this expansion into \eqref{crazy} for the Bremsstrahlung function we obtain
\begin{equation}
\label{higherorder}
B_{1/2}(\lambda) = \frac{\lambda }{8}-\frac{\pi ^2 }{48}\lambda ^3 +\frac{\pi ^4}{60}\lambda ^5-\frac{841 \pi ^6 }{40320}\lambda ^7+\frac{2963 \pi ^8 }{90720}\lambda ^9+\frac{196959097 \pi ^{10}}{159667200}\lambda ^{11} + {\cal O}\left(\lambda ^{12}\right)
\end{equation}

\end{document}